\documentclass[11pt]{article}

\usepackage{amsmath}
\usepackage{graphicx,psfrag,epsf}
\usepackage{enumerate}
\usepackage{natbib}
\usepackage{url} % not crucial - just used below for the URL

\usepackage{authblk}
\usepackage{subfig}
\usepackage{amssymb,amsthm}
\usepackage{wrapfig}
\usepackage{multirow}
\let\chapter\section % Fix for algo2e
\usepackage[titlenumbered,linesnumbered,ruled,vlined]{algorithm2e}
\usepackage{caption}
\usepackage{bm}
\usepackage{booktabs}
\usepackage[smallscripts]{moresize}
\newcommand{\blind}{0}

\usepackage{hyperref}           % For hyperlinks and indexing the PDF
\hypersetup{ % play with the different link colors here
    colorlinks,
    citecolor=blue,
    filecolor=blue,
    linkcolor=blue,
    urlcolor=blue % set to black to prevent printing blue links
}

\newcommand{\bu}{\mathbf{u}}
\newcommand{\bv}{\mathbf{v}}
\newcommand{\bx}{\mathbf{x}}
\newcommand{\by}{\mathbf{y}}

\newcommand{\bC}{\mathbf{C}}
\newcommand{\bE}{\mathbf{E}}

\newcommand{\bP}{\mathbf{P}}
\newcommand{\bQ}{\mathbf{Q}}
\newcommand{\bS}{\mathbf{S}}
\newcommand{\bR}{\mathbf{R}}
\newcommand{\bU}{\mathbf{U}}
\newcommand{\bV}{\mathbf{V}}

\newcommand{\bX}{\mathbf{X}}
\newcommand{\bY}{\mathbf{Y}}

\newcommand{\bvep}{\bm{\varepsilon}}

\newcommand{\hv}{\widehat \bv}
\newcommand{\hu}{\widehat \bu}
\newcommand{\hC}{\widehat \bC}

\newcommand{\supp}{\mathrm{supp}}

\newcommand{\tp}{^{T}}

\newcommand{\loss} {\mathcal{L}}

\DeclareMathOperator*{\argmin}{argmin}

\newtheorem{theorem}{Theorem}

\newtheorem{condition}{Condition}

\newtheorem{corollary}{Corollary}
\newtheorem{propose}{Proposition}
\newcommand{\eat}[1]{}

\begin{document}

\def\spacingset#1{\renewcommand{\baselinestretch}%
{#1}\small\normalsize} \spacingset{1}

%%%%%%%%%%%%%%%%%%%%%%%%%%%%%%%%%%%%%%%%%%%%%%%%%%%%%%%%%%%%%%%%%%%%%%%%%%%%%%

%\if1\blind
%{
%  \title{\bf RASE: Scalable Sparse Reduced-Rank Regression}
%  \author{Mohammad Taha Bahadori\thanks{
%    Mohammad Taha Bahadori (E-mail: \textit{bahadori@gatech.edu}) is a Postdoctoral Fellow in School of Computational Science and Engineering, Georgia Tech, Atlanta, GA 30332. This work was done when he was a PhD student at University of Southern California, and his work was supported by NSF award number 1254206.}\hspace{.2cm}\\
%    School of Computational Science and Engineering, Georgia Tech\\
%    and \\
%    Sanjay Purushotham \\
%    Department of ZZZ, University of WWW}
%  \maketitle
%} \fi
%
%\if0\blind
%{
%  \bigskip
%  \bigskip
%  \bigskip
%  \begin{center}
%    {\LARGE\bf RASE: Scalable Sparse Reduced-Rank Regression}
%\end{center}
%  \medskip
%} \fi

\if0\blind
{
\title{%\textbf{Scalable Estimation in Sparse Reduced-Rank Regression with Application to Social Media Analysis}
%RASE: Scalable Sparse Reduced-Rank Regression
%Scalable sparse reduced-rank regression via the SEED
%Scalable Interpretable Multi-Response Regression via the SEED
Scalable Interpretable Multi-Response Regression via SEED
\thanks{Mohammad Taha Bahadori (E-mail: \textit{bahadori@gatech.edu}) is Postdoctoral Fellow, School of Computational Science and Engineering, Georgia Institute of Technology, Atlanta, GA 30332. %This work was done when he was a PhD student at University of Southern California. %and his work was supported by NSF award number 1254206.
Zemin Zheng is Associate Professor (E-mail: \textit{zhengzm@ustc.edu.cn}), Department of Statistics and Finance, School of Management, University of Science and Technology of China, China. %This work was supported by Start-up Grant by USTC.
%Emre Demirkaya is Ph.D. Candidate (E-mail: \textit{demirkay@usc.edu}), Department of Mathematics, University of Southern California, Los Angeles, CA 90089.
%Sanjay Purushotham is Postdoctoral Scholar (Email: \textit{spurusho@usc.edu}), Computer Science Department, Viterbi School of Engineering, University of Southern California, Los Angeles, CA 90089.
Yan Liu is Associate Professor (E-mail: \textit{yanliu.cs@usc.edu}), Computer Science Department, Viterbi School of Engineering, University of Southern California, Los Angeles, CA 90089.
%This work was supported by NSF award number 1254206 and Okawa Foundation Research Award.
Jinchi Lv is McAlister Associate Professor in Business Administration (E-mail: \textit{jinchilv@marshall.usc.edu}), Data Sciences and Operations Department, Marshall School of Business, University of Southern California, Los Angeles, CA 90089.
%This work was supported by NSF CAREER Award DMS-0955316.
This work was supported by NSF CAREER Awards DMS-0955316 and IIS-1254206, Okawa Foundation Research Award, and a grant from the Simons Foundation. The authors would like to thank Emre Demirkaya and Sanjay Purushotham for their help with discussions and proofreading.
}%
\date{\today}
%\date{July 9, 2014}
\author{%Mohammad Taha Bahadori,
M. Taha Bahadori, Zemin Zheng, Yan Liu and Jinchi Lv
%\medskip\\
%University of Southern California\\
} %
}
} \fi

\if1\blind
{
  \bigskip
  \bigskip
  \bigskip
\title{%\textbf{Scalable Estimation in Sparse Reduced-Rank Regression with Application to Social Media Analysis}
%RASE: Scalable Sparse Reduced-Rank Regression%
Scalable Interpretable Multi-Response Regression via SEED
}
  \medskip
} \fi

\maketitle

\spacingset{1.45}

\begin{abstract}
Sparse reduced-rank regression is an important tool to uncover meaningful dependence structure between large numbers of predictors and responses in many big data applications such as genome-wide association studies and social media analysis. Despite the recent theoretical and algorithmic advances, scalable estimation of sparse reduced-rank regression remains largely unexplored. In this paper, we suggest a scalable procedure called sequential estimation with eigen-decomposition (SEED) which needs only a single top-$r$ singular value decomposition to find the optimal low-rank and sparse matrix by solving a sparse generalized eigenvalue problem. Our suggested method is not only scalable but also performs simultaneous dimensionality reduction and variable selection. Under some mild regularity conditions, we show that SEED enjoys nice sampling properties including consistency in estimation, rank selection, prediction, and model selection. Numerical studies on synthetic and real data sets show that SEED outperforms the state-of-the-art approaches for large-scale matrix estimation problem.
\end{abstract}

\textit{Running title}: SEED %Scalable Sparse Reduced-Rank Regression

\textit{Key words}: Reduced-rank regression; Scalability; Sparsity; High dimensionality; Greedy algorithm;  %Greedy variable selection;
Sparse eigenvector estimation %; Social media analysis

%\newpage
\section{Introduction}
\label{sec:intro}

Identifying complex dependence structures among predictors and responses is an important problem in statistics and machine learning, since these structures reveal hidden domain knowledge about the data. For example, in bioinformatics, identifying gene regulatory networks is crucial for understanding gene regulatory paths and gene functions, which helps disease prediction and diagnosis. Similarly, in social media analysis, inferring the influence networks from user activities (that is, \textit{Diffusion Network Inference} problem \citep{Leskovec2010,Zhou2013,embar2014bayesian}) is an important problem, and it has found applications in social media marketing \citep{gomez2012} and crisis management \citep{starbird2012will}. In these big data applications, inferring the dependence structures is challenging since the responses and predictors may be related through a few latent pathways and/or associated through only a subset of responses and predictors. Moreover, the curse of dimensionality and massive amounts of data, that is, scalability issues make the dependence structure discovery problem even harder to solve. Recently,  regularization methods such as lasso \citep{lasso} and group lasso \citep{grouplasso}, and reduced-rank regression approaches \citep{velu2013multivariate,izenman1975reduced} have been proposed to recover sparse response-predictor associations and latent predictors, respectively. \cite{chen2015note} and \cite{chen2012reduced} have proposed sparse reduced-rank regression approaches by combining the regularization and reduced-rank regression techniques to find the complex dependence structures between responses and predictors.

Sparse reduced-rank regression works by modeling the associations between the predictor and response variables via a sparse and low-rank representation of the coefficient matrix. It not only enhances the interpretability of the estimated matrix by eliminating irrelevant features \citep{chen2012reduced}, but also reduces the number of free parameters of the model and thus the number of observations required for desired estimation consistency \citep{Negahban2011,Yuan2007,bunea2011,Candes2011,Chen2013}. Sparse reduced-rank regression has found applications in micro-array biclustering \citep{chen2012reduced}, subspace clustering \citep{wang2013provable}, social network community discovery \citep{richard2014link, Zhou2013}, and motion segmentation \citep{feng2014robust}. In these applications, joint sparsity and low-rankness has been used to enforce a clustered dependence structure among data points.  In particular, the key idea is to estimate a similarity matrix among data points that is simultaneously sparse and low-rank and then permute the rows and columns of the matrix to yield approximately block-diagonal structures, which naturally lead to clustering of data points into several groups. Note that \cite{agarwal2012noisy}, \cite{chandrasekaran2010latent}, and the references therein have considered estimating matrices with a low-rank plus sparse representation which is different from our work as we are interested in estimating a matrix that is jointly low-rank and sparse.

A natural approach to solving the sparse reduced-rank regression problem is to simultaneously penalize the parameter matrix using the $L_1$ and nuclear norm regularizers, as they are convex relaxations to sparsity and low-rankness of a matrix, respectively. The resulting optimization problem is convex and can be solved using the alternating direction method of multipliers (ADMM) \citep{Boyd_2010} as proposed by \cite{richard2014link} and \cite{Zhou2013}. In \cite{bunea2012joint}, an alternative approach, called rank constrained group lasso (RCGL), was proposed which directly penalizes the rank and the number of nonzero rows of the parameter matrix. They showed oracle rates for the estimated matrix and also provided a practical algorithm which iteratively and jointly solves a $L_1$-regularization and low-rank estimation problem. To further improve the estimation accuracy, \cite{chen2012reduced} borrowed ideas from adaptive Lasso \citep{zou2006adaptive} and proposed the iterative exclusive extraction algorithm (IEEA) which finds a locally optimal solution in the neighborhood of the initial value. They also showed model selection consistency and asymptotic normality results along with the improved empirical performance of IEEA on microarray biclustering analysis data.

All the above approaches for sparse reduced-rank regression achieve both desirable theoretical properties and strong empirical results. However, they cannot scale to large matrix estimation problems in many big data applications. The ADMM algorithm of \cite{richard2014link} and \cite{Zhou2013} uses iterative singular value thresholding \citep{cai2010singular} for solving the joint $L_1$ and nuclear norm regularization. Iterative singular value thresholding is known to be computationally expensive since it performs a full singular value decomposition of the parameter matrix in each iteration. On the other hand, RCGL \citep{bunea2012joint} is computationally much faster than ADMM since it only performs top-$r$ singular value decomposition for estimating a rank-$r$ matrix in each iteration. Despite a lower computational cost per iteration, it is unclear how many iterations RCGL needs for convergence. IEEA \citep{chen2012reduced} performs nested loops of alternating $L_1$-penalized regression for each singular vector which can be expensive, especially on parallel computing devices. The iterative nature of these three approaches makes them not scalable and renders them inefficient for large matrix estimation even on high performance computing devices.

To overcome the scalability issues of the previous approaches, we propose a simple and scalable sparse reduced-rank regression procedure called sequential estimation with eigen-decomposition (SEED). SEED is designed for high-performing computing platforms. It converts the sparse and low-rank regression problem to a sparse generalized eigenvalue problem, and then solves the problem using the recent algorithms for sparse eigenvalue decomposition \citep{Yuan2013, Ma2013, cai2013}. As a pure learning algorithm, SEED is expected to perform only a single top-$r$ sparse eigenvalue decomposition for estimating a rank-$r$ matrix, which makes it truly scalable and efficient for large matrix estimation problems.

The main contributions of our paper are threefold. First, for the sparse reduced-rank regression problem, our proposed procedure SEED provides a scalable approach to uncovering the sparse predictor-response association network while simultaneously achieving dimension reduction and variable selection. Second, for the high-dimensional settings, our theoretical analysis shows that SEED can consistently estimate the singular vectors, latent factors as well as the regression coefficient matrix, identify the correct rank of the matrix, accurately predict the multivariate response vector, and recover the support of the singular vectors under mild conditions. Note that, compared to \cite{chen2012reduced}, we do not make any assumption on the positive definiteness of the design matrix for proving our consistency results. %In comparison with \cite{chen2012reduced}, while we provide the similar $\mathcal{O}(n^{-1/2})$ parameter estimation consistency rate in high-dimensions, our results do not rely on the assumption of positive definiteness  of the design matrix.
Third, we empirically demonstrate that SEED can not only be efficiently implemented on both central processing units (CPU) and graphics processing units (GPU) for large-scale applications, but it also outperforms the state-of-the-art sparse reduced-rank regression approaches.

The rest of this paper is organized as follows. Section \ref{sec:method} introduces our SEED method. We discuss the implementation details of SEED in Section \ref{sec:imp} and present its asymptotic properties in Section \ref{sec:analysis}. We demonstrate the advantages of SEED on both synthetic and real data sets  in Section \ref{sec:exp}, and in Section \ref{sec:dis} we discuss some extensions of our SEED method. All technical proofs and details are provided in the Appendix.

\section{Sequential estimation with eigen-decomposition (SEED)}
\label{sec:method}

\subsection{Data model and problem formulation}\label{sec2.1}
%%Let $\{(\mathbf{x}_i, \mathbf{y}_i)\}_{i = 1}^{n}$ denote a sample of $n$ independent and identically distributed (i.i.d.) observations $(\mathbf{x}_i, \mathbf{y}_i)$ where $\mathbf{x}_i \in \mathbb{R}^{p}$ and $\mathbf{y}_i \in \mathbb{R}^{q}$ represent the $i$th predictor  and the corresponding response vectors, respectively. Given a predictor vector $\mathbf{x}$, the corresponding response vector $\mathbf{y}$ is drawn from the following model
Denote by $\{(\mathbf{x}_i, \mathbf{y}_i)\}_{i = 1}^{n}$ $n$ observations for the fixed design setting where $\mathbf{x}_i \in \mathbb{R}^{p}$ and $\mathbf{y}_i \in \mathbb{R}^{q}$ represent the $i$th predictor  and the corresponding response vectors, respectively. Given a predictor vector $\mathbf{x}$, the corresponding response vector $\mathbf{y}$ is drawn from the following model
\begin{equation*}
\mathbf{y} = \bC^{*T}\mathbf{x} + \bm{\varepsilon},
%\label{eq:model}
\end{equation*}
where the noise vector $\bm{\varepsilon} \sim \mathcal{N}(\mathbf{0}, \bm{\Sigma})$ is a $q$-dimensional zero mean multivariate Gaussian random vector with the covariance matrix $\bm{\Sigma}$, and $\bC^* \in \mathbb{R}^{p\times q}$ is the regression coefficient matrix \footnote{Note that the Gaussianity of noise variables is not essential to either our procedure or the theoretical results and similar results would hold under the sub-Gaussian assumption \cite[Chapter 14]{buhlmann2011statistics}.}. We can rewrite the data model in the matrix form as follows
\begin{equation}
\bY = \bX\bC^* + \bE,
\label{eq:model}
\end{equation}
\noindent where $\bY = [\by_1, \ldots, \by_n]\tp $, $\bX = [\bx_1, \ldots, \bx_n]\tp$, and $\bE = [\bvep_1, \ldots, \bvep_n ]\tp $ denote the matrices of stacked response, predictor and noise vectors, respectively.

%%For simplicity, we study the case of fixed design setting.
Let $\bP = n^{-1}\bX^{T}\bX$ be the Gram matrix of the predictors. We consider the case where the true regression coefficient matrix $\bC^*$ is jointly low-rank and sparse, the same as in \cite{bunea2012joint} and \cite{chen2012reduced}. In particular, the matrix rank $r^*$ is assumed to be small with $r^* \ll \min(p, q)$, and $\bC^*$ follows a decomposition that
\begin{equation}
\bC^* = \sum_{k=1}^{r^*} \mathbf{u}_k^*\mathbf{v}_k^{*T} = \sum_{k=1}^{r^*} \bC_k^*,
\label{eq:C*}
\end{equation}
where the left singular vectors $\bu_k^* \in \mathbb{R}^{p}$ are $\bP$-orthogonal with unit length, that is, $\bu_k^{*T}\bP\bu_{k'}^* = 0$ if $k\neq k'$ and $\|\bu_k^*\|_2 = 1$, while the right singular vectors $\bv_k^* \in \mathbb{R}^{q}$ are orthogonal, that is, $\bv_k^{*T}\bv_{k'}^* = 0$ for $k\neq k'$, and $\bC_k^*$ is the layer $k$ unit rank matrix of $\bC^*$. The singular vectors are sorted by the magnitudes of the singular values $\sigma_k = \frac{1}{\sqrt{q n}}\|\bX\bC_k^*\|_F$ in descending order, which correspond to their contributions to the prediction of $\bY$.

We consider the left singular vectors (both the population and estimated ones) in the constraint space $\bu \perp \mathrm{Ker}(\bP)$, where $\mathrm{Ker}(\bP)$ denotes the null space of $\bP$, to guarantee the model identifiability as otherwise $\bu$ would contain certain component $\widetilde{\bu}$ such that $\bX \widetilde{\bu} = 0$, which does not contribute to the prediction of $\bY$. The $\bP$-orthogonality of $\bu_k^*$ is not necessary in the algorithm but facilitates the theoretical analysis. In fact, the above decomposition (\ref{eq:C*}) for $\bC^*$ coincides with the singular value decomposition of $\bX \bC^*$ through different scalings on the singular vectors. We defer the discussion on the existence and identifiability of decomposition (\ref{eq:C*}) to Supplementary Material \ref{sec:ident}.

The aforementioned modeling of the regression coefficient matrix indeed gives a latent factor model with $r^*$ latent factors, where $\bX\bu_k^*$ is the $k$th latent factor and $\bv_k^*$ describes the impact of the $k$th factor on the response variables. As illustrated in \cite{Yuan2007}, the low-rankness of $\bC^*$ renders dimension reduction such that all responses can be predicted by a relatively small set of common factors. Furthermore, the left singular vectors $\bu_k^*$ correspond to the selection of predictors, and we impose a sparsity assumption such that $\left\| \mathbf{u}_k^* \right\|_0 \ll p$ for $k = 1, \ldots, r^*$. Similar sparsity assumptions were made in \cite{bunea2012joint} and \cite{chen2012reduced} to enhance model interpretability by removing irrelevant features in high dimensions, where \cite{chen2012reduced} assumed that both the left and right singular vectors are sparse while \cite{bunea2012joint} imposed restriction on the number of nonzero rows of the coefficient matrix. In this paper, we are interested in two cases: (i) when the right singular vectors are not required to be sparse and (ii) when it is desirable to have sparse right singular vectors, which entails the response selection. We will show that both cases are efficiently accommodated by our procedure.

Our goal is to accurately estimate the singular vectors $\mathbf{u}_k^*$ and $\mathbf{v}_k^*$, and the true rank $r^*$ such that we can recover the latent factors as well as their impacts, and at the same time, identify the underlying number of latent factors and the significant predictors. As a singular vector can have two opposite directions, we always assume that the estimated left singular vector takes the correct one, that is, the angles between estimated and population left singular vectors are no more than a right angle. Once the estimated rank and singular vectors are obtained, the estimate $\widehat{\bC}$ of the true matrix $\bC^*$ follows immediately from (\ref{eq:C*}). For theoretical analysis, we consider the estimation of the true rank within certain upper bound $r > r^*$, that is, $\mathrm{rank}(\widehat{\bC}) \leq r$. In practice, the upper bound $r$ can be controlled by our algorithm. Unlike most existing sparse and low-rank estimation methods which adopt the regularization framework of a loss function plus certain penalties and are generally not scalable, the proposed procedure SEED is indeed a pure learning algorithm that predicts $\bY$ using $\bX \widehat{\bC}$ with some low-rank $\widehat{\bC}$.

%For the compactness of notation, we use $S_r$ to denote the set of all matrices with rank no larger than $r$ and satisfy the aforementioned sparsity constraint, and for any matrix $\bC$, the Frobenius norm is defined as $\|\bC\|_F^2 = \sum_{i,j} \bC_{i, j}^2$. We propose to solve the following reduced-rank multi-task regression problem to obtain $\hC$:
%\begin{align}
%\hC &= \argmin_{\bC \in S_r}\left\|\bY - \bX\bC\right\|_F^2.
%\label{eq:theProb}
%\end{align}
%(A measure of fitting, not actual objective function).

\subsection{Description of SEED}
%In order to design a scalable method for sparse reduced-rank regression, we adapt the greedy variable selection framework. The fundamental principle of greedy variable selection is finding the greedy direction that results in the largest decrease in the loss function. In greedy sparse regression problems the dictionary has a finite size, i.e., as many as the number of covariates.  Thus, the greedy procedures can search over the entire dictionary and find the optimal direction \cite{Tropp2007,Barron2008,Zhang2011}.  In contrast, there are infinitely many rank-$1$ matrices which makes the search very challenging. In this section, we first describe how we can find the optimal rank-1 matrix, then we present our proposed procedure RASE.
The following proposition provides us insight into estimating the top-$r^*$ left and right singular vectors of $\bC^*$.

\begin{propose}
\label{lem:rank1opt}
Consider the noiseless case where $\bY^* = \bX\bC^*$ and $\bC^* = \sum_{k=1}^{r^*}\mathbf{u}_k^*\mathbf{v}_k^{*T}$ as defined in (\ref{eq:C*}). Then $\mathbf{u}_1^*, \ldots, \mathbf{u}_{r^*}^*$ are the $r^*$ non-degenerate left singular vectors of $\bC^*$ if and only if they are the eigenvectors of the following generalized eigenvalue problem
\begin{equation}
\bX\tp \bY^*\bY^{*T} \bX\mathbf{u} = \lambda\bX\tp\bX\mathbf{u}
\label{eq:geneig}
\end{equation}
with respect to the nonzero eigenvalues $\lambda_1, \cdots, \lambda_{r^*}$, where $\lambda_k = nq \sigma_k^2$ is the $k$th largest eigenvalue of $\bY^*\bY^{*T}$ with the singular values $\sigma_k$ defined in Section \ref{sec2.1}. Furthermore, given the left singular vector $\mathbf{u}_k^*$, the corresponding right singular vector $\mathbf{v}_k^*$ can be written as
\begin{align}\label{eq:solu1}
\mathbf{v}_k^* = \frac{1}{\mathbf{u}_k^{*T}\bX\tp\bX\mathbf{u}_k^*}\bY^{*T} \bX\mathbf{u}_k^*.
\end{align}
%On the other hand, when $\bX$ is of full rank (either full row rank or full column rank), constrained on the space $\bu \perp \mathrm{Ker}(\bP)$, the generalized eigenvalue problem $(\ref{eq:geneig})$ has exactly $r^*$ nonzero eigenvalues $\lambda_1, \cdots, \lambda_{r^*}$ with the corresponding eigenvectors $\mathbf{u}_1^*, \cdots, \mathbf{u}_{r^*}^*$.
\end{propose}
%A detailed proof of Proposition \ref{lem:rank1opt} is provided in Appendix \ref{sec:optProof}.
Proposition \ref{lem:rank1opt} shows that the problem of estimating the singular vectors can be transformed into the generalized eigenvalue problem in (\ref{eq:geneig}), thanks to the $\bP$-orthogonality of the left singular vectors. In the noisy case where $\bY = \bX\bC^* + \bE$, it motivates us to estimate the left singular vectors by solving the corresponding generalized eigenvalue problem as follows
\begin{equation}
\bX\tp \bY\bY^{T} \bX\mathbf{u} = \lambda\bX\tp\bX\mathbf{u}.
\label{eq:geneig1}
\end{equation}
The estimation consistency in the noisy case will be guaranteed by the matrix perturbation theory, see Section \ref{sec:analysis} for details. On the other hand, it is not difficult to see that the eigenvectors with respect to different eigenvalues of problem (\ref{eq:geneig1}) are $\bP$-orthogonal, which further gives the orthogonality of the right singular vectors estimated by (\ref{eq:solu1}). It implies that the right and left singular vectors obtained by solving the generalized eigenvalue problem (\ref{eq:geneig1}) will automatically be orthogonal and $\bP$-orthogonal, respectively.

Related results of principal component analysis in low dimensions can be found in \cite{baldi1989neural}, \cite{de2003framework}, and \cite{diamantaras1996principal}. Note that in the high-dimensional setting, the regime of interest for this paper, the Gram matrix $\bP$ can be rank deficient and the generalized eigenvalue problem is potentially challenging to solve. We will address the implementation challenges for high-dimensional settings in Section \ref{sec:imp}.
%Moreover, as pointed out in \cite{bunea2012joint}, when the predictor matrix $\bX$ is not of full column rank, we can always reduce $\bX$ to a matrix with $\mathrm{rank}(\bX)$ independent columns that span the same space as the columns of $\bX$.
%Note that when $\bY$ has an additive noise term, the solution for $\mathbf{v}_k$ in the proposition neither uses the fact that the right singular vectors should be orthogonal, nor guarantees their orthogonality, however we will show that they are close to be orthogonal.

Motivated by Proposition \ref{lem:rank1opt}, our proposed procedure SEED performs a two-step estimation for the regression coefficient matrix: it first solves the generalized eigenvalue problem in (\ref{eq:geneig1}) to obtain the estimated left singular vectors $\hu_1, \ldots, \hu_r$ with unit length; then it finds the estimated right singular vectors $\hv_1, \ldots, \hv_r$ according to (\ref{eq:solu1}), that is,
\begin{align}
&\widehat{\mathbf{v}}_k = \frac{1}{\widehat{\mathbf{u}}_k^{T}\bX\tp\bX\widehat{\mathbf{u}}_k}\bY\tp\bX\widehat{\mathbf{u}}_k. \label{eq:solu}
\end{align}
The maximum rank $r$ depends on the magnitude of the estimated singular value $\widehat{\sigma}_k = \frac{1}{\sqrt{q n}}\|\bX\widehat{\bC}_k\|_F$ with $\widehat{\bC}_k = \widehat{\mathbf{u}}_k \widehat{\mathbf{v}}_k^T$ (whether it is larger than a threshold $\mu$). And the optimal rank will be tuned by the information criterion described in Section \ref{sec:analysis}.

The details of the procedure are provided in Algorithm \ref{algo:all}. To achieve a sparse solution, we need to find the optimal rank-$r$ sparse matrix via a sparse eigenvalue decomposition procedure in Line \ref{line:eig} of \ref{algo:all}. For theoretical analysis, we assume that there exists a sparse eigenvalue decomposition procedure that solves (\ref{eq:geneig1}) and practical methods will be provided in Section \ref{sec:imp}. The practical methods need a sparsity parameter $\theta$, such as a threshold \citep{Ma2013} or a sparsity size \citep{Yuan2013}. We will show in Section \ref{sec:exp} that SEED is robust to the choices of parameters $\theta$ and $\mu$. If the right singular vectors are also required to be sparse, we perform a simple element-wise thresholding on $\widehat{\mathbf{v}}_k$ after we obtain it in Line \ref{line:th}.

\begin{algorithm}[t]
 \SetAlgoLined
 \DontPrintSemicolon
 \KwIn{$\bY \in \mathbb{R}^{n\times q}$ and $\bX \in \mathbb{R}^{n\times p}$.}
 \KwIn{A termination accuracy $\mu$ and sparsity parameter $\theta$.}
 Compute matrices:  $\bP \gets \frac{1}{n}\bX\tp\bX,\; \bR \gets \frac{1}{n}\bY\tp\bX, \; \bQ \gets \frac{1}{q}\bR^{\top}\bR$, \;
 $\widehat{\bC} \gets \mathbf{0}_{q\times p}$\;
 $k \gets 1$\;
 \Repeat{$\widehat{\sigma}_k < \mu$}{
 $(\widehat{\mathbf{u}}_{k}, \widehat{\sigma}_k) \gets k$th $\theta$-sparse eigenvector and eigenvalue of $\bQ\mathbf{u} = \lambda \bP\mathbf{u}$.\label{line:eig}\;
% $\delta_k \gets (\widehat{\mathbf{v}}_{k}^{\top}Q\widehat{\mathbf{v}}_{k})/(\widehat{\mathbf{v}}_{k}^{\top}P\widehat{\mathbf{v}}_{k})$.\;
 \If{$\widehat{\sigma}_k > \mu$}{
 $\widehat{\mathbf{v}}_{k} \gets \frac{1}{\widehat{\mathbf{u}}_{k}^{\top}\bP\widehat{\mathbf{u}}_{k}}\bR\widehat{\mathbf{u}}_{k},\qquad$ \textit{\#Optional thresholding of $\widehat{\mathbf{v}}_{k}$ for sparsity}.\label{line:th} \;
 $\widehat{\bC} \gets \widehat{\bC} + \widehat{\mathbf{u}}_{k}\widehat{\mathbf{v}}_{k}^{\top}$.\;
 $k \gets k + 1$.\,
 }
 }
  \Return{$\widehat{\bC}$}\;
\caption{}
%\TitleOfAlgo{SEED}
 \label{algo:all}
\end{algorithm}
%\begin{equation*}
%\fbox{\text{Algorithm SEED should be about here.}}
%\end{equation*}

Given Condition \ref{as:RSC} in Section \ref{sec:analysis}, the following proposition shows that with significant probability, \ref{algo:all} will stop after exactly $r^*$ iterations if the termination criterion $\mu$ is properly chosen.
\begin{propose} Suppose that Condition \ref{as:RSC} holds and $\log(pq) = o(n)$, then with probability at least $1-\delta$ for any $\delta \in (0, 1)$, \ref{algo:all} will stop after $r^*$ iterations if the termination criterion $\mu$ is chosen from the following interval:
\begin{equation*}
C\sqrt{\frac{1}{n}\log \frac{pq}{\delta}} \leq \mu \leq \sigma_{r^*}^2 - C\sqrt{\frac{1}{n}\log \frac{pq}{\delta}},
\end{equation*}
\noindent where $\sigma_{r^*}^2$ is assumed to be larger than $2C\sqrt{\frac{1}{n}\log \frac{pq}{\delta}}$ with positive constant $C$ defined in Supplementary Material \ref{sec:rankproof}.
\label{prop:termin}
\end{propose}
Proposition \ref{prop:termin} is mainly for theoretical purposes since the interval above is generally unknown to us. We will need to tune the optimal rank by certain information criterion in practice. The tail probability $\delta$ can decay to zero quickly as $p$ and $q$ grow with rates such as $\delta \propto (pq)^{-\alpha}$ for some positive constant $\alpha > 1$. This is due to the fact that when $\delta \propto (pq)^{-\alpha}$, we have $\sqrt{\frac{1}{n}\log \frac{pq}{\delta}} \to 0$ under the assumption that $\log(pq) = o(n)$. The lower bound of $\mu$ excludes the case that extra latent factors are involved due to noises while the upper bound guarantees the important factors will not be missed.

\section{Scalable implementation of SEED}
\label{sec:imp}

\ref{algo:all} requires a sparse solution of (\ref{eq:geneig1}) which is a generalized eigenvalue problem with a rank deficient matrix $\bP$. In this section, we study multiple practical aspects of SEED and propose two different ways to solve (\ref{eq:geneig1}), one with enhanced stability and the other using a fast procedure to accelerate it.

\subsection{Basic implementation}
\paragraph{Stability.} For numerical stability purposes, we can solve the following modified problem of (\ref{eq:geneig1}) with a very small positive $\rho$:
\begin{equation}
\bX\tp \bY\bY^{T} \bX\mathbf{u} = \lambda (\bX\tp\bX + \rho \mathbf{I})\mathbf{u}.
\label{eq:geneig2}
\end{equation}
Note that $\bX\tp\bX + \rho \mathbf{I}$ is invertible since the eigenvalues of $\bX\tp\bX$ are nonnegative. Denote by $\widetilde{\bX} \in \mathbb{R}^{p\times p}$ the modified predictor matrix such that $\widetilde{\bX}^{T}\widetilde{\bX} = \bX\tp\bX + \rho \mathbf{I}$, which can be obtained via the Cholesky decomposition, and $\widetilde{\bY} = (\widetilde{\bX}\tp)^{-1}\bX\tp\bY$ the modified response matrix. Then the above equation (\ref{eq:geneig2}) can be rewritten as
\begin{equation}
\widetilde{\bX}\tp \widetilde{\bY}\widetilde{\bY}^{T} \widetilde{\bX}\mathbf{u} = \lambda \widetilde{\bX}\tp\widetilde{\bX}\mathbf{u},
\label{eq:regu3}
\end{equation}
which adopts the same form as (\ref{eq:geneig1}).

The formulation of $\widetilde{\bX}$ and $\widetilde{\bY}$ can be regarded as a generalization of the ridge regression to the multivariate response setting. In fact, since $\bC^*$ is the minimizer of $\left\|\bY^* - \bX\bC \right\|_F^2$, we can enhance the stability by adding a small Frobenius norm regularization as follows:
\begin{equation*}
\widetilde{\bC} = \argmin_{\bC}\left\{\left\| \bY - \bX\bC\right\|_F^2 + \rho \left\| \bC\right\|_F^2\right\},
\end{equation*}
%$S_r$ denotes the set of all matrices with rank no larger than $r$ and satisfy the aforementioned sparsity constraint
where the Frobenius norm is defined as $\|\bC\|_F^2 = \sum_{i,j} \bC_{i, j}^2$ for any matrix $\bC$. After completing the squares, we get
\begin{equation*}
\widetilde{\bC} = \argmin_{\bC}\left\{\| \widetilde{\bY} - \widetilde{\bX}\bC\|_F^2\right\},
\end{equation*}
which means that $\widetilde{\bX}$ and $\widetilde{\bY}$ are the corresponding predictor and response matrices that take into account the shrinkage effects \citep{stein,shrinkage}.

A computationally efficient technique for solving equation (\ref{eq:regu3}) is to solve the sparse eigenvalue problem $\widetilde{\bP}^{-1}\widetilde{\bQ}\mathbf{u} = \lambda \mathbf{u}$, where the modified Gram matrix $\widetilde{\bP} = n^{-1}\widetilde{\bX}\tp\widetilde{\bX}$ and $\widetilde{\bQ}$ is defined accordingly as in Algorithm \ref{algo:all}. We also use the Sherman--Morrison--Woodbury formula to compute $\widetilde{\bP}^{-1}$ as follows:
\begin{equation*}
(\rho \mathbf{I}_p + \bX\tp\bX)^{-1} = \frac{1}{\rho}\mathbf{I}_p - \frac{1}{\rho^2}\bX\tp(\mathbf{I}_n + \frac{1}{\rho}\bX\bX\tp)^{-1}\bX.
\end{equation*}
The above equation requires inversion of an $n\times n$ matrix instead of a $p\times p$ matrix which is significantly faster in the high-dimensional setting when $p \gg n$. The corresponding right singular vectors will then be obtained by (\ref{eq:solu}).
%such that they are orthogonal to each other.
%Another advantage of this approach is that both sets of estimated singular vectors are orthogonal.

\paragraph{Refitting.} In \ref{algo:all}, a further refitting can be performed during the eigenvalue decomposition in Line \ref{line:eig} to enhance the stability. The refitting procedure is as follows. In the $k$th step, we compute the residual $\bY_k = \bY - \bX \sum_{j = 1}^{k - 1}\hC_j$ and solve the generalized eigenvalue problem (\ref{eq:geneig1}) with $\bY$ replaced by $\bY_k$ to obtained the unit rank matrix $\hC_k = \hu_k \hv_k^T$. Then we perform the top-$k$ singular value decomposition $\hC = \bU\widehat{\bS}\bV^{T}$ for $\hC = \sum_{j = 1}^{k}\hC_j$ and refit the solution by finding $\widetilde{\bS} = \argmin_{\bS}\left\|\bY-\bX\bU\bS\bV^{T} \right\|_F^2$. The estimate with refitting is defined as $\widetilde{\bC} = \bU\widetilde{\bS}\bV^{T}$. In practice, we find this approach more stable and report the results based on this variation of SEED in numerical studies.

\subsection{Scalability}
\paragraph{Speedup.} The bottleneck in speeding up Algorithm \ref{algo:all} is Line \ref{line:eig} where we need to solve a sparse generalized eigenvalue problem. To overcome this bottleneck, we propose a new solution to estimating the left singular vectors by rewriting equation (\ref{eq:geneig1}) as
\begin{equation*}
\bX\tp(\bY\bY\tp - \lambda \mathbf{I}) \bX\mathbf{u} = \mathbf{0}.
%\label{eq:lambdaMax}
\end{equation*}
%Eq. (\ref{eq:lambdaMax}) shows that the maximum value of $\lambda$ cannot exceed the largest eigenvalue of $\bY\bY\tp$ denoted by $\lambda_{\max}(\bY\bY\tp)$. Furthermore, when $\mathrm{rank}(\bX)= n$, $\bX\mathbf{u}$ can span the entire $n$-dimensional Euclidean space; thus given the dominant eigenvector $\bu_0$ for the problem in Eq. (\ref{eq:lambdaMax}), $\bX\bu_0$ will be the dominant eigenvector for $\bY\bY\tp$.
Similar to Proposition \ref{lem:rank1opt}, when $\bX$ is of full row rank (which is easy to satisfy in the high-dimensional setting), the above equation shares the same nonzero eigenvalues with $\bY\bY\tp$. Even if $\bX$ is row rank deficient, the nonzero solution of $\bu$ is ensured by the perturbation theory in Lemma \ref{lem:pert}. Thus, we propose the following two-step procedure for Line \ref{line:eig} of \ref{algo:all}:
\begin{itemize}
\item[(1)] $\lambda \gets \lambda_{\max}(\bY\bY\tp)$.
\item[(2)] $\widehat{\mathbf{u}} \gets \text{sparse eigenvector corresponding to zero eigenvalue of }(\bX\tp\bY\bY\tp\bX - \lambda \bX\tp\bX )$.
\end{itemize}
%If there is no nonzero solution in the second step above (which may happen when $\bX$ is row rank deficient), it means $\lambda_k$ is only an eigenvalue of $\bY\bY\tp$ but not the generalized eigenvalue problem (\ref{eq:geneig1}). Then we just skip $\lambda_k$ and turn to the next one.
Similarly as before, we compute the residual $\bY_k = \bY - \bX \sum_{j = 1}^{k - 1}\hC_j$ in the $k$th step and replace $\bY$ with $\bY_k$ in the above two-step procedure to obtain the $k$th left singular vector $\hu_k$. Overall, the first step requires calculation of the top-$r$ eigenvalues for an $n\times n$ matrix (or $q\times q$ if $q < n$) while the second step finds the corresponding eigenvectors by solving a regular sparse eigenvalue problem. Thus, the above procedure significantly accelerates the speed of SEED as it converts a degenerate sparse generalized eigenvalue problem to two simpler regular sparse eigenvalue problems. The truncated power method \citep{Yuan2013} can be used to compute both eigenvalue problems efficiently.
%Note that this method is only an approximation for scalability purposes and may fail in pathological cases. In the cases where $\mathrm{rank}(\bX) < n$, the above procedure is an approximation because $\bX\mathbf{u}$ may not span the entire $n$-dimensional space.

\paragraph{Sparse eigenvector estimation.}
The previous two approaches for solving the generalized eigenvalue decomposition problem indicate that we can solve the problem via regular eigenvalue decomposition. This allows us to reuse the existing procedures for sparse eigenvalue decomposition such as \cite{Yuan2013}, \cite{Ma2013}, \cite{cai2013}, and \cite{lei2015} to solve the problem in (\ref{eq:geneig1}). In numerical studies, we use the iterative thresholding method \citep{Ma2013} for estimating the sparse eigenvectors of a matrix.

%This property enables faster tuning compared to the regularization based algorithms where we need to solve a separate optimization algorithm for every regularization parameter. \emph{[TB] Is this true?}

\paragraph{Parallel implementation.} In order to scale up the procedures, we often need to utilize the parallel computing tools. Given the fact that SEED only uses basic matrix operations, we can employ parallel implementation of the large matrix operations to accelerate SEED. In Section \ref{sec:exp}, our experiments with GPU which contain thousands of processing units show that the matrix operations in SEED can be efficiently parallelized and it significantly enhances the speed of SEED. Whenever the data can not be loaded into the memory of a single device, efficient distributed algorithms can be used, see for example, \cite{Kang2011} and the references therein.

\section{Asymptotic properties of SEED}
\label{sec:analysis}
%\input{analysis.tex}

%%%%%%%%%%%%%%%%%%%%%%%%%%%%%%%%%%%%%%%%%%%%%%%%%%%%%%%%%%%%%%%%%%%%%%%%%%%%%%%%%%%%%%%%%%%
In this section, we will analyze the statistical properties of SEED. Define the maximum sparsity level of the left singular vectors as $s^* = \max_{k = 1}^{r^*} \|\mathbf{u}_k^*\|_0 \ll p$, which is assumed mainly for theoretical analysis (see Condition \ref{as:RSC} below) and will not be used directly in the algorithm. We consider the estimated left singular vectors with the number of nonzero elements less than certain sparsity level $s > s^*$, that is, $\|\widehat{\bu}_k\|_0 \leq s$ for $k = 1, \cdots, r$.
%To facilitate the theoretical analysis, we assume that the predictor matrix $\bX$ is of full rank, which is kind of reasonable in high-dimensional linear regression setting (\ref{eq:model}) since some observation can be redundant if it is a linear combination of the others in the noiseless case.
%We also denote the largest element of the noise covariance matrix $\bm{\Sigma}^*$ by $\gamma^2 \triangleq \max_{j,k}\bm{\Sigma}_{jk}^*$.
%Whenever there is an ambiguity, we will denote the true parameter with a star superscript.
%As defined in Proposition \ref{lem:rank1opt}, the magnitudes of right singular vectors satisfy $\frac{1}{\sqrt{q}}\|\mathbf{v}_k^*\|_2 = a_k$, while the left singular vectors are assumed to be $\bP$-orthogonal and have unit norm $\|\mathbf{u}_k^{*}\|_2 = 1$. For ease of presentation, $A = \max_{k = 1}^{r^*} a_k$ is regarded as a constant.
%It can be relaxed by adding a Frobenius norm regularization, as described in the previous section.

%In order to analyze accuracy of RASE, we need to compare its solution with the optimal noiseless solution.  The optimal noiseless solution is given by:
%\begin{align}
%\bC^* &= \argmin_{\bC,~ \bC \in S_{r^*}}\left\|\mathbb{E}[\bY|\bX] - \bX\bC\right\|_F^2, \label{eq:rankkoptimal}
%\end{align}
%\noindent whereas RASE solves the problem in Eq. (\ref{eq:theProb}).
%\eat{
%\begin{align}
%\hC &= \argmin_{\bC,~ \bC \in S_{r^*}}\left\|\bY - \bX\bC\right\|_F^2. \label{eq:rankkestimate}
%\end{align}
%}

\subsection{Technical conditions}
\label{sec:conds}
%Our first goal is to bound the error in estimation of $\bC_k^*$ by $\hC_k$. We need the following assumptions:
%We first list our assumptions; note that not all of the assumptions are required for establishing individual theorems.
Here we list a few technical conditions and discuss their relevance in detail.

\begin{condition} [Restricted isometry] There exists a positive constant $\phi_s$ such that the Gram matrix $\bP$ satisfies
\[\phi_s \leq \min_{\mathbf{z}\in \mathbb{R}^{p}}\left\{\frac{\|\bP\mathbf{z}\|_2}{\|\mathbf{z}\|_2}: \|\mathbf{z}\|_0 \leq 2s \right\} \leq \max_{\mathbf{z}\in \mathbb{R}^{p}}\left\{\frac{\|\bP\mathbf{z}\|_2}{\|\mathbf{z}\|_2}: \|\mathbf{z}\|_0 \leq 2s \right\} \leq \phi_s^{-1}\]
for some $s > s^*$.
\label{as:RSC}
\end{condition}

\begin{condition} [Minimum singular value separation] The non-zero singular values $\sigma_k$ satisfy $\sigma_{k}^2 - \sigma_{k + 1}^2 \geq d_{\sigma} > 0$ for some constant $d_{\sigma}$ and $k = 1, \ldots, r^* - 1$.
\label{as:separtion}
\end{condition}

\begin{condition} [Bounded eigenvalues]
The eigenvalues of the population covariance matrix of the noise vector $\bm{\varepsilon}$ satisfy $0 < \gamma_l^2 \leq \lambda_j(\bm{\Sigma}) \leq \gamma_u^2 < \infty$ for $j = 1, \ldots, q$, where $\gamma_l$ and $\gamma_u$ are positive constants with $\gamma_u \leq c_{\gamma} \sigma_{r^*}$ for some positive constant $c_{\gamma}$.
\label{as:lowerBound}
\end{condition}

\begin{condition} [Minimum signal strength]
\label{as:supp}
There exists some positive constant $\delta \in(0, 1)$ such that the following lower bounds on the magnitudes of the non-zero elements of $\mathbf{u}_k^*$ and $\mathbf{v}_k^*$ hold for any $1 \leq k \leq r^*$:
\begin{align*}
\min_{i\in \supp(\mathbf{u}_k^*)} |u_i^*| &\geq 3C_u \sqrt{\frac{1}{n}\log \frac{pq}{\delta}},\\
\min_{i\in \supp(\mathbf{v}_k^*)} \frac{1}{\sqrt{q}}|v_i^*| &\geq 3C_v \sqrt{\frac{1}{n}\log \frac{pq}{\delta}},
\end{align*}
where $C_u$ and $C_v$ are constants defined in Theorem \ref{th:rankm}.
\end{condition}

Condition \ref{as:RSC} imposes bounds on the $2s$-sparse eigenvalues of $\bP$, which is weaker than the regular bounded eigenvalue assumption since the sparse eigenvalues do not grow as fast as the regular eigenvalues when the dimensionality $p$ grows. As a typical condition in high dimensions, it restricts the correlations between small numbers of features and thus guarantees the identifiability of the true sparse support. See, for instance, \cite{candes2005} and \cite{Zhang2011} for more discussion on it.
%It is similar to the unit-norm normalization in sparse regression problem, e.g. \cite[Assumption 3.1]{Zhang2011}.

Recall that $\sigma_k$ is the $k$th largest singular value of $\frac{1}{\sqrt{qn}}\bX\bC^*$. Condition \ref{as:separtion} requires a non-zero separation among the singular values such that the true left singular vectors are distinguishable. For ease of presentation, we assume $d_{\sigma}$ to be a constant. In fact, $d_{\sigma}$ can be allowed to converge to zero asymptotically, so we indicate the effect of $d_{\sigma}$ clearly in the constants of the theoretical results.

The elements of the unobserved noise vector $\bm{\varepsilon}$ were assumed to be independent and identically distributed (i.i.d.) in \cite{bunea2012joint}. We relax it a bit in Condition \ref{as:lowerBound} by imposing bounded eigenvalues for the noise covariance matrix to recover the true rank in Theorem \ref{th:rankconsist}. Our technical argument still applies when either $\gamma_l \to 0$ or $\gamma_u \to \infty$ as long as their rates of convergence can be controlled within certain magnitudes.

The two inequalities in Condition \ref{as:supp} are imposed for the model selection consistency of the predictors and responses, respectively. The magnitude of the minimum signal strength is $O\left(\sqrt{\frac{\log (pq)}{n}}\right)$, which is relatively mild as it converges to zero in our setting. Since $\bu_k^*$ is assumed to have unit length, the scaling of $\bC^*$ is put on $\bv_k^*$ such that there is an extra factor $\frac{1}{\sqrt{q}}$ in the second inequality.

%%%%%%%%%%%%%%%%%%%%%%%%%%%%%%%%%%%%%%%%%%%%%%%%%%%%%%%%%%%%%%%%%%%%%%%%%%%%%%%%%%%%%%%%%%%
%%%%%%%%%%%%%%%%%%%%%%%%%%%%%%%%%%%%%%%%%%%%%%%%%%%%%%%%%%%%%%%%%%%%%%%%%%%%%%%%%%%%%%%%%%%
\subsection{Main results}
Denote by $P^2 = \max_{j = 1}^p \bP_{jj}$ and $\gamma^2 = \max_{j = 1}^q \bm{\Sigma}_{jj}$ the maximum diagonal component of the Gram matrix and noise covariance matrix, respectively. Without loss of generality, we assume that $V = \max_{k = 1}^{r^*} \frac{1}{\sqrt{q}}\|\mathbf{v}_k^{*}\|_2$ is finite. Moreover, it is clear that under Conditions \ref{as:RSC} and \ref{as:lowerBound}, $P$ and $\gamma$ are also finite constants. Throughout this section, we assume that the top-$r$ eigenvectors are obtained by solving the generalized eigenvalue problem (\ref{eq:geneig1}) with the maximum sparsity level $s$ such that Condition \ref{as:RSC} is satisfied. The estimated regression coefficient matrix is given by $\hC = \sum_{k=1}^{\tilde{r}}\widehat{\bC}_k$, where $\tilde{r}$ is the optimal rank tuned by information criterion (\ref{inform}) and $\hC_k =  \widehat{\mathbf{u}}_k\widehat{\mathbf{v}}_k^{T}$. The following theorem bounds the estimation errors of SEED with the estimated left singular vectors $\hu_k$ taking the correct signs as discussed before.
\begin{theorem} [Estimation and prediction consistency]
\label{th:rankm}
Suppose that Conditions \ref{as:RSC} and \ref{as:separtion} hold, $\gamma$ is finite, and $\log (pq) = o(n)$, then with probability at least $1 - \delta$ for any $\delta \in (0, 1)$ and uniformly over $k = 1, \ldots, r^*$, we have
\begin{align*}
\|\widehat{\mathbf{u}}_k - \mathbf{u}^*_k \|_2 & \leq C_u \sqrt{\frac{1}{n}\log \frac{pq}{\delta}} + o\left(\sqrt{\frac{1}{n}\log \frac{pq}{\delta}} \ \right),\\
\frac{1}{\sqrt{q}}\|\widehat{\mathbf{v}}_k - \mathbf{v}^{*}_k\|_{2} & \leq C_v\sqrt{\frac{1}{n}\log \frac{pq}{\delta}} + o\left(\sqrt{\frac{1}{n}\log \frac{pq}{\delta}} \ \right),\\
\frac{1}{\sqrt{n}}\left\|\bX( \widehat{\mathbf{u}}_k - \mathbf{u}_k^*)\right\|_2 &\leq \phi^{-1/2} C_u\sqrt{\frac{1}{n}\log \frac{pq}{\delta}} +o\left(\sqrt{\frac{1}{n}\log \frac{pq}{\delta}}\right),\\
\frac{1}{\sqrt{q}}\|\widehat{\bC}_k - \bC^*_k\|_F &\leq (V C_u + C_v) \left(\sqrt{\frac{1}{n}\log \frac{pq}{\delta}} \ \right) + o\left(\sqrt{\frac{1}{n}\log \frac{pq}{\delta}} \ \right),\\
\frac{1}{\sqrt{qn}} \|\bX(\widehat{\bC}_k - \bC^*_k)\|_F &\leq \phi_s^{-1/2} (V C_u + C_v) \left(\sqrt{\frac{1}{n}\log \frac{pq}{\delta}} \ \right) + o\left(\sqrt{\frac{1}{n}\log \frac{pq}{\delta}} \ \right),
\end{align*}
where the constants $C_u =  \frac{4 \gamma P \sigma_1}{d_{\sigma}\phi_{s}^{5/2}}$ and $C_v = 2\sqrt{2}\phi_s^{-3/2}(2V \phi_s^{-1/2} + \sigma_1)C_u + 2\sqrt{2}\phi_s^{-1} \gamma P$.
\end{theorem}

Theorem \ref{th:rankm} shows that the uniform estimation error bounds for both top-$r^*$ singular vectors $\mathbf{u}^*_k$ and $\frac{1}{\sqrt{q}} \mathbf{v}^*_k$, the top-$r^*$ latent factors $\frac{1}{\sqrt{n}}\bX\bu^*_k$ and unit rank matrices $\frac{1}{\sqrt{q}} \bC^*_k$, and the uniform prediction error bounds of the top-$r^*$ latent factors are all in the same order of $O\left(\sqrt{\frac{1}{n}\log \frac{pq}{\delta}} \ \right)$. Similar to Proposition \ref{prop:termin}, by setting $\delta = (pq)^{-\alpha}$ with $\alpha > 1$, the tail probability will decay to zero quickly as the dimensionality $p$ and $q$ grow. Furthermore, the estimation and prediction accuracy would then be within the rate of $O\left(\sqrt{\frac{\log (pq)}{n}} \ \right)$, where the factor $\log (pq)$ reflects the curse of dimensionality as there are $pq$ parameters in total from the regression coefficient matrix $\bC^*$.

If the true rank $r^*$ can be correctly identified, it is not difficult to see that the estimation accuracy for $\frac{1}{\sqrt{q}}\bC^*$ will be within the rate of $O\left(\sqrt{ \frac{r^* \log (pq)}{n}} \ \right)$ (see Corollary \ref{corr1} below), which coincides with the minimax error bound for estimating the regression coefficient vector in the univariate response setting \citep{yu2011} with the dimensionality $p$ and sparsity size $s$ replaced by the overall dimensionality $pq$ and true rank $r^*$, respectively. In view of this, the true rank $r^*$, instead of the maximum sparsity level $s^*$, plays the same role in multivariate regression as the sparsity size in the univariate response setup. With the true rank $r^*$, the prediction accuracy for the multivariate response vector $\by$ will also follow from the prediction accuracy of the top-$r^*$ latent factors.

%%%%%%%%%%%%%%%%%%%%%%%%%%%%%%%%%%%%%%%%%%%%%%%%%%%%%%%%%%%%%%%%%%%%%%%%%%%%%%%%%%%%%%%%%%%
%%%%%%%%%%%%%%%%%%%%%%%%%%%%%%%%%%%%%%%%%%%%%%%%%%%%%%%%%%%%%%%%%%%%%%%%%%%%%%%%%%%%%%%%%%%
Based on the discussion before, a desirable statistical property of any low-rank estimation procedure is to accurately recover the true rank of the parameter matrix. Similar to Lasso, the nuclear norm regularization needs to be enhanced by techniques such as adaptive regularization to accurately recover the rank of the matrix \citep{Chen2013,bach2008consistency}.
In contrast, in SEED we can directly control the rank of the solution by limiting the number of steps. In particular, we propose a GIC-type \citep{Fan2013} information criterion that guarantees rank recovery by SEED when the optimal rank is tuned according to it.

%In fact, tuning RASE is more intuitive and flexible than the regularization approaches where we need to tune the regularization parameters. The following theorem identifies a range for $\mu$ in Algorithm \ref{algo:all} to guarantee rank consistency of RASE.

%Given the following true model for the data:
%
%\begin{equation}
%Y = \sum_{i=1}^{r}A_iX + E.
%\end{equation}
%\noindent where $A_i$ are sorted according to $\sigma_i = \frac{1}{n}\|A_iX\|_2$.

\begin{theorem} [Consistency of rank recovery]
Suppose Conditions \ref{as:RSC}--\ref{as:lowerBound} hold, $\log (pq) = o(n)$, $r = o \left(\left[\frac{n}{\log (pq)}\right]^{1/4}\right)$, and $r^* = o \left(\frac{1}{\sqrt{\log \log n}} \cdot \left[\frac{n}{\log (pq)}\right]^{1/4}\right)$. Define the following information criterion:
\begin{equation}\label{inform}
\mathcal{C}_n = \sqrt{n}\log\loss_n(\bY, \bX, \hC) + \mathrm{rank}(\hC)\sqrt{\log(pq)} \log \log n,
\end{equation}
\noindent where $\loss_n(\bY, \bX, \bC) = \frac{1}{qn}\|\bY-\bX\bC\|_F^2$. Under the above information criterion, with probability at least $1 - (pq)^{-\alpha}$ for some positive constant $\alpha > 1$ and sufficiently large $n$, SEED will select the true rank, that is, $\mathrm{rank}(\hC) = \mathrm{rank}(\bC^*)$.
\label{th:rankconsist}
\end{theorem}
%The proof is provided in Appendix \ref{sec:rankproof}.
%%%%%%%%%%%%%%%%%%%%%%%%%%%%%%%%%%%%%%%%%%%%%%%%%%%%%%%%%%%%%%%%%%%%%%%%%%%%%%%%%%%%%%%%%%%
%%%%%%%%%%%%%%%%%%%%%%%%%%%%%%%%%%%%%%%%%%%%%%%%%%%%%%%%%%%%%%%%%%%%%%%%%%%%%%%%%%%%%%%%%%%

In the high-dimensional setting where the number of predictors can increase exponentially with the sample size, it is demonstrated in \cite{Fan2013} that we need some power of the logarithmic factor of dimensionality ($\sqrt{\log (pq)}$ for our setting) in the model complexity penalty of the information criterion to consistently identify the true model, and the slow diverging rate $\log \log n$ is set to prevent underfitting. The proof of Theorem \ref{th:rankconsist} indeed shows that information criterion (\ref{inform}) will keep decreasing until the estimated rank reaches the true rank $r^*$, where in each step the amount of decrease in the objective function $\loss_n(\bY, \bX, \hC)$ equals to the squared singular value obtained by solving the generalized eigenvalue problem (\ref{eq:geneig1}). After reaching the true rank, the estimated singular value becomes small such that the model complexity penalty will overweight the decrease and then information criterion (\ref{inform}) would start increasing. Therefore, in the sequence of solutions generated by SEED, the estimate $\hC$ with rank $r^*$ will be the minimizer of (\ref{inform}) such that the true rank can be correctly identified.

As discussed before, correct identification of the true rank will yield the estimation accuracy of $\bC^*$ as well as the prediction accuracy of $\bX \bC^*$. Therefore, combined with Theorem \ref{th:rankconsist}, it is immediate that the results in Theorem \ref{th:rankm} give the following corollary.
\begin{corollary} [Overall estimation and prediction consistency] Given Conditions \ref{as:RSC}--\ref{as:lowerBound}, $\log (pq) = o(n)$, $r = o \left(\left[\frac{n}{\log (pq)}\right]^{1/4}\right)$, and $r^* = o \left(\frac{1}{\sqrt{\log \log n}} \cdot \left[\frac{n}{\log (pq)}\right]^{1/4}\right)$, if the optimal rank is tuned by information criterion (\ref{inform}), then with probability at least $1 - (pq)^{-\alpha}$ for any constant $\alpha > 0$ and sufficiently large $n$, we have
\begin{align*}
\frac{1}{\sqrt{q}}\|\widehat{\bC} - \bC^*\|_F &\leq (1 + \alpha)(V C_u + C_v) \left(\sqrt{ \frac{r^* \log (pq)}{n}} \ \right) + o\left(\sqrt{ \frac{r^* \log (pq)}{n}} \ \right),\\
\frac{1}{\sqrt{qn}} \|\bX(\widehat{\bC} - \bC^*)\|_F &\leq \phi_s^{-1/2} (1 + \alpha) (V C_u + C_v) \left(\sqrt{ \frac{r^* \log (pq)}{n}} \ \right) + o\left(\sqrt{ \frac{r^* \log (pq)}{n}} \ \right).
\end{align*}
\label{corr1}
\end{corollary}

Besides estimation consistency and rank recovery, SEED is also able to find the true support of the singular vectors. To achieve this goal, after selecting the optimal rank, we need to further refine the model selection procedure by performing a hard-thresholding. See, for example, \cite{FanLv2013} for more general implications of thresholding in high-dimensional sparse modeling. Specifically, denote by $T_{\theta}(\mathbf{z})$ the estimator after the hard-thresholding operation on every element of $\mathbf{z} = (z_1, \cdots, z_p) \in \mathbb{R}^{p}$ such that
\begin{equation*}
T_{\theta}(z_{i}) = \left\{ \begin{array}{lr} 0 & \text{if} ~|z_{i}| < \theta\\
z_{i}& \text{otherwise}\end{array} \right., \quad i = 1, \ldots, p.
\end{equation*}
%We need the following condition:
Based on the results of Theorems \ref{th:rankm} and \ref{th:rankconsist} and the signal strength assumption in Condition \ref{as:supp}, we have the following properties for the estimator with a further thresholding.
\begin{theorem} [Support recovery of the singular vectors] Given Conditions \ref{as:RSC}--\ref{as:supp}, $\log (pq) = o(n)$, $r = o \left(\left[\frac{n}{\log (pq)}\right]^{1/4}\right)$, and $r^* = o \left(\frac{1}{\sqrt{\log \log n}} \cdot \left[\frac{n}{\log (pq)}\right]^{1/4}\right)$, for every pair of singular vectors, $(\hu_k, \hv_k)$, $k = 1, \ldots, r^*$, the following results hold:
\begin{itemize}
\item[a)] If the threshold $\theta \in (\frac{5}{4} T_u, \frac{7}{4} T_u)$ with $T_u = C_u \sqrt{\frac{1}{n}\log \frac{pq}{\delta}}$, then with probability at least $1 - \delta$, we have $\supp(T_{\theta}(\widehat{\mathbf{u}}_k)) = \supp(\mathbf{u}_k^*)$;
\item[b)] If the threshold $\theta \in (\frac{5}{4} T_v, \frac{7}{4} T_v)$ with $T_v = C_v \sqrt{\frac{1}{n}\log \frac{pq}{\delta}}$, then with probability at least $1 - \delta$, we have $\supp(T_{\theta}(\widehat{\mathbf{v}}_k)) = \supp(\mathbf{v}_k^*)$.
\end{itemize}
\label{cor:supp}
\end{theorem}
%The proof is given in Appendix \ref{sec:suppproof}.

Theorem \ref{cor:supp} shows that both supports of the left and right singular vectors can be accurately recovered with properly chosen tuning parameter $\theta$. Together with the correctly identified true rank $r^*$, the above results indeed yield  consistent selection of both predictors and responses. In practice, this threshold $\theta$ can be tuned by criteria such as cross-validation. Our simulation studies in Section \ref{sec:exp} show that SEED is robust to the choice of tuning parameter $\theta$.

%\paragraph{Conclusion.} %Corollary \ref{cor:supp} indicates that the high-dimensionality  shows its effects only on the right singular vector $\mathbf{v}$ and the support of the left singular vector $\mathbf{u}$ can be efficiently recovered by simple thresholding. This observation has been made also in \cite{bunea2012joint} where the authors use $L_{2, 1}$ norm to impose sparsity only on the columns of the solution.  Note that in our approach can simply obtain column sparse solution by not thresholding the left singular vectors which is computationally more efficient that $L_{2, 1}$ norm regularization approach.
Besides the statistical properties established before, the proposed procedure SEED enjoys great flexibility in the sense that it does not rely on exact eigenvalue decomposition and the perturbation errors in the generalized eigenvalue problem (\ref{eq:geneig1}) will be linearly incorporated into the estimated singular vectors $\widehat{\mathbf{u}}_k$ and $\widehat{\mathbf{v}}_k$. Furthermore, our analysis does not reply on the positive definiteness of the Gram matrix \citep{chen2012reduced} in high dimensions.

\section{Numerical studies}
\label{sec:exp}

In this section, we conduct experiments on three data sets, including two simulation data sets (one for a medium-scale experiment and one for a large-scale experiment) and one application data set in social media analysis, to examine the empirical performance of SEED.  %The medium-scale experiments are intended to compare the accuracy of RASE with the baselines. The large scale experiments are conducted on high performance machines to show the power of RASE in processing enormous problems.  The application experiments show the  of the proposed algorithm in real-world data sets.

\subsection{Simulation studies}

\subsubsection{Simulation example 1}
We generate a medium-scale synthetic data set as follows: the predictors $\mathbf{x}$ are drawn from a multivariate Gaussian distribution as $\mathbf{x} \sim \mathcal{N}(\mathbf{0}_{p\times 1}, \bm{\Sigma}_X)$, where $\bm{\Sigma}_X$ is the $p\times p$ covariance matrix with auto-regressive structure, that is, $\Sigma_{X,i,j} = \rho^{|i-j|}$ for some $0 <\rho < 1$ which will be specified later.  The responses $\mathbf{y}$ are drawn according to conditional distribution $\mathbf{y}|\mathbf{x} \sim \mathcal{N}(\bC^{\top}\mathbf{x}, \gamma \bm{\Sigma}_E)$ where the noise covariance matrix $\bm{\Sigma}_E$ is also selected to have the autoregressive structure with $\rho = 0.5$ and we set $\gamma = 0.1$. We generate the parameter matrix $\bC$ as follows: first we generate a block-sparse matrix $\widetilde{\bC}$  with $5\%$ non-zero elements. Each non-zero element of $\widetilde{\bC}$ is drawn from a $\mathcal{N}(0, 1)$.  To achieve a low-rank structure, we find the top-$r$ singular value decomposition of $\widetilde{\bC}$ as $\widetilde{\bC} = \bU\bS\bV\tp$, and then set the elements of $\bU$ and $\bV$ whose magnitude is smaller than $0.01$ to zero to obtain $\bar{\bU}$ and $\bar{\bV}$. The final parameter matrix is obtained as $\bC = \bar{\bU}\bar{\bS}\bar{\bV}\tp$ where the first $r$ diagonal elements of the diagonal matrix $\bar{\bS}$ are set to $100, 99, \ldots, 101-r$. Without loss of generality, we add a few vectors to the design matrix to ensure the orthogonality condition of Section \ref{sec:conds}. In all of the simulation experiments, we generate 100 data sets and report the mean and standard error of performance for different methods.
%and randomly set only $5\%$ of the elements of $\mathbf{u}$ and $\mathbf{v}$ to $1$ to achieve sparseness. Then, in order to satisfy the assumption on orthogonality of $\mathbf{u}_k$ to the kernel of $\bX$, we append $\bX$ with few more vectors.

We compare the performance of SEED with two state-of-art methods: (1) RCGL \citep{bunea2012joint} and (2) Penalized regression with simultaneous $L_1$ and nuclear-norm penalization. The optimization problem is solved by the popular alternating direction method of multipliers \citep{Boyd_2010} and we will refer to this baseline as the ``LN--ADMM'' algorithm. %Details of the algorithm are provided in Appendix \ref{sec:admm}.
All model parameters are set based on a separate validation set with size $n_{\text{valid}} = 500$.% when the dimension ($p$) is high.

The quality of the estimator $\widehat{\bC}$ is evaluated via four performance metrics listed as follows. (1) \textit{Normalized Prediction Error} defined as:
\begin{equation*}
\mathrm{Normalized~Prediction~Error}  = \frac{\|\bY_{\mathrm{test}} - \bX_{\mathrm{test}}\widehat{\bC}\|_F}{\|\bY_{\mathrm{test}}\|_F}.
\end{equation*}
(2) \textit{Normalized Parameter Estimation Error} defined as:
\begin{equation*}
\mathrm{Normalized~ Parameter ~Estimation~Error} = \frac{\|\widehat{\bC} - \bC \|_F}{\|\bC\|_F},
\end{equation*}
where $\bC$ is the true parameter matrix.

\noindent(3) \textit{Rank Recovery Error} defined as:
\begin{equation*}
\mathrm{Rank~Recovery~Error} = |\mathrm{rank}(\widehat{\bC}) - \mathrm{rank}(\bC)|.
\end{equation*}
Since the solution of the nuclear norm always leads to small non-zero singular values (which prevents $\widehat{\bC}$ from being low-rank), we threshold the singular values of $\widehat{\bC}$ that are more than 100 times smaller than its largest singular value to have a fair comparison.

(4) \noindent\textit{Support Recovery AUC}, that is, the area under the receiver operating characteristic (ROC) curve of comparing support of $\widehat{\bC}$ with the ground truth, which is always  between 0 and 1. It is computed by varying the decision threshold and obtaining the false positive and true positive curve. Then the area under the false positive and true positive curve is reported as AUC. The value of AUC indicates the probability that a procedure assigns a higher value to a randomly chosen non-zero element than a randomly chosen zero element \citep{hanley1982meaning}. It is an appropriate metric for measuring support recovery accuracy because in sparse support recovery we have more zeros than non-zeros which inflates the result of the simple 0-1 accuracy measure. In contrast, AUC is more robust to imbalanced positive/negative prediction labels.

Table \ref{tab:synth} shows the results of all algorithms on a variety of regimes by varying the dimensionality $p$ and the rank $r$.  We can see that SEED is superior or comparable to the baseline algorithms across all  four measures. As the results show and the theory predicts, in most high-dimensional cases, nuclear norm usually overestimates the true rank of the matrix. Furthermore, we find that the iterative SVD procedure in the RCGL algorithm often results in significant underestimation of the true rank, when the true rank is large. Note that in addition to accuracy, SEED also significantly reduces the variance of the estimation.
%%\begin{equation*}
%%\fbox{\text{Table \ref{tab:synth} should be about here.}}
%%\end{equation*}

\begin{table}%[t]
	\caption{Simulation Results}
	\centering \scriptsize
	\begin{tabular}{@{}l|c | c |c| c|c }
		\toprule
		\multirow{2}{*}{$p$} & Algorithm & Normalized Prediction  & Normalized Estimation  & Rank Recovery  & Support Recovery\\
		& & Error ($\times 10^{-2}$) & Error ($\times 10^{-2}$) &Error &  AUC\\
		\toprule
		\toprule
		\multicolumn{6}{c}{$n = 100$, $q = 200$, $\mathbf{r = 3}$, and $\rho = 0.5$}\\ \toprule
		\multirow{3}{*}{$100$}
		& SEED & $\bm{1.8958 ~(0.0587)}$ & $\bm{0.5762 ~(0.0312)}$ & $\bm{0.0000 ~(0.0000)}$ & $\bm{0.9816~(0.0026)}$\\
		& LN--ADMM & $3.1031 ~(0.0529)$ & $12.4132 ~(0.1736)$ & $0.6100 ~(0.0634)$ & $0.8078 ~(0.0020)$\\
		& RCGL  & $4.3006 ~(0.0619)$ & $16.3268 ~(0.1708)$ & $0.7600 ~(0.0698)$ & $0.8024 ~(0.0019)$\\
		\midrule
		\multirow{3}{*}{$400$}
		& SEED & $\bm{0.8140 ~(0.0121)}$ & $\bm{0.3568 ~(0.0069)}$ & $\bm{0.0000 ~(0.0000)}$ & $\bm{0.9706~(0.0017)}$\\
		& LN--ADMM & $13.5766 ~(0.1356)$ & $27.3787 ~(0.1328)$ & $1.6700 ~(0.0711)$ & $0.8085 ~(0.0050)$\\
		& RCGL  & $29.7221 ~(0.2244)$ & $34.6811 ~(0.2674)$ & $0.8400 ~(0.0692)$ & $0.7174 ~(0.0029)$\\
		\midrule
		\multirow{3}{*}{$800$}
		& SEED & $\bm{0.6296 ~(0.0084)}$ & $\bm{0.3603 ~(0.0063)}$ & $\bm{0.0000 ~(0.0000)}$ & $\bm{0.9720~(0.0012)}$\\
		& LN--ADMM & $5.6560 ~(0.0504)$ & $15.8111 ~(0.0774)$ & $0.3500 ~(0.0500)$ & $0.9948 ~(0.0001)$\\
		& RCGL  & $6.5201 ~(0.0419)$ & $19.4005 ~(0.0847)$ & $0.4700 ~(0.0627)$ & $0.9911 ~(0.0001)$\\
		\midrule
		\multirow{3}{*}{$1500$}
		& SEED & $\bm{0.4891 ~(0.0048)}$ & $\bm{0.3578 ~(0.0052)}$ & $\bm{0.0000 ~(0.0000)}$ & $\bm{0.9718~(0.0011)}$\\
		& LN--ADMM & $5.4634 ~(0.0238)$ & $22.6509 ~(0.0533)$ & $3.0700 ~(0.0624)$ & $0.9949 ~(0.0001)$\\
		& RCGL  & $8.2969 ~(0.0481)$ & $29.3042 ~(0.0780)$ & $1.3700 ~(0.2159)$ & $0.9871 ~(0.0001)$\\
		\midrule
		\multirow{3}{*}{$2000$}
		& SEED & $\bm{0.4351 ~(0.0049)}$ & $\bm{0.3546 ~(0.0060)}$ & $\bm{0.0000 ~(0.0000)}$ & $\bm{0.9738~(0.0009)}$\\
		& LN--ADMM & $4.9001 ~(0.0224)$ & $23.1360 ~(0.0656)$ & $2.8700 ~(0.0812)$ & $0.9962 ~(0.0001)$\\
		& RCGL  & $8.7672 ~(0.0501)$ & $32.0000 ~(0.0897)$ & $2.5100 ~(0.3043)$ & $0.9879 ~(0.0001)$\\
		\bottomrule
		\multicolumn{6}{c}{$n = 100$, $q = 200$, $\mathbf{r = 30}$, and $\rho = 0.5$}\\
		%\toprule
		%$p$ & Algorithm & Normalized Prediction Error & Normalized Estimation Error & Support Recovery AUC & Rank Recovery Error\\
		\toprule
		\multirow{3}{*}{$100$}
		& SEED & $\bm{0.5820 ~(0.0058)}$ & $\bm{2.2812 ~(0.0144)}$ & $\bm{0.0000 ~(0.0000)}$ & $\bm{0.8132~(0.0022)}$\\
		& LN--ADMM & $3.1031 ~(0.0529)$ & $12.4132 ~(0.1736)$ & $0.6100 ~(0.0634)$ & $0.8078 ~(0.0020)$\\
		& RCGL  & $4.3006 ~(0.0619)$ & $16.3268 ~(0.1708)$ & $0.7600 ~(0.0698)$ & $0.8024 ~(0.0019)$\\
		\midrule
		\multirow{3}{*}{$400$}
		& SEED & $\bm{1.6704 ~(0.1641)}$ & $\bm{7.1831 ~(0.6961)}$ & $\bm{0.0000 ~(0.0000)}$ & $\bm{0.8053~(0.0044)}$\\
		& LN--ADMM & $13.5766 ~(0.1356)$ & $27.3787 ~(0.1328)$ & $1.6700 ~(0.0711)$ & $\bm{0.8085 ~(0.0050)}$\\
		& RCGL  & $29.7221 ~(0.2244)$ & $34.6811 ~(0.2674)$ & $0.8400 ~(0.0692)$ & $0.7174 ~(0.0029)$\\
		\midrule
		\multirow{3}{*}{$800$}
		& SEED & $\bm{0.1721 ~(0.0014)}$ & $\bm{0.4592 ~(0.0046)}$ & $\bm{0.0000 ~(0.0000)}$ & $\bm{0.9993~(0.0000)}$\\
		& LN--ADMM & $5.6560 ~(0.0504)$ & $15.8111 ~(0.0774)$ & $0.3500 ~(0.0500)$ & $0.9948 ~(0.0001)$\\
		& RCGL  & $6.5201 ~(0.0419)$ & $19.4005 ~(0.0847)$ & $0.4700 ~(0.0627)$ & $0.9911 ~(0.0001)$\\
		\midrule
		\multirow{3}{*}{$1500$}
		& SEED & $\bm{0.1299 ~(0.0008)}$ & $\bm{0.4440 ~(0.0031)}$ & $\bm{0.0000 ~(0.0000)}$ & $\bm{0.9992~(0.0000)}$\\
		& LN--ADMM & $5.4634 ~(0.0238)$ & $22.6509 ~(0.0533)$ & $3.0700 ~(0.0624)$ & $0.9949 ~(0.0001)$\\
		& RCGL  & $8.2969 ~(0.0481)$ & $29.3042 ~(0.0780)$ & $1.3700 ~(0.2159)$ & $0.9871 ~(0.0001)$\\
		\midrule
		\multirow{3}{*}{$2000$}
		& SEED & $\bm{0.1130 ~(0.0007)}$ & $\bm{0.4380 ~(0.0032)}$ & $\bm{0.0000 ~(0.0000)}$ & $\bm{0.9992~(0.0000)}$\\
		& LN--ADMM & $4.9001 ~(0.0224)$ & $23.1360 ~(0.0656)$ & $2.8700 ~(0.0812)$ & $0.9962 ~(0.0001)$\\
		& RCGL  & $8.7672 ~(0.0501)$ & $32.0000 ~(0.0897)$ & $2.5100 ~(0.3043)$ & $0.9879 ~(0.0001)$\\
		\bottomrule
	\end{tabular}
	\label{tab:synth}
\end{table}
%\end{landscape}

\begin{figure}[t!]
	\centering
	\subfloat[Rank]{\label{fig:rankPath}\includegraphics[scale=0.25]{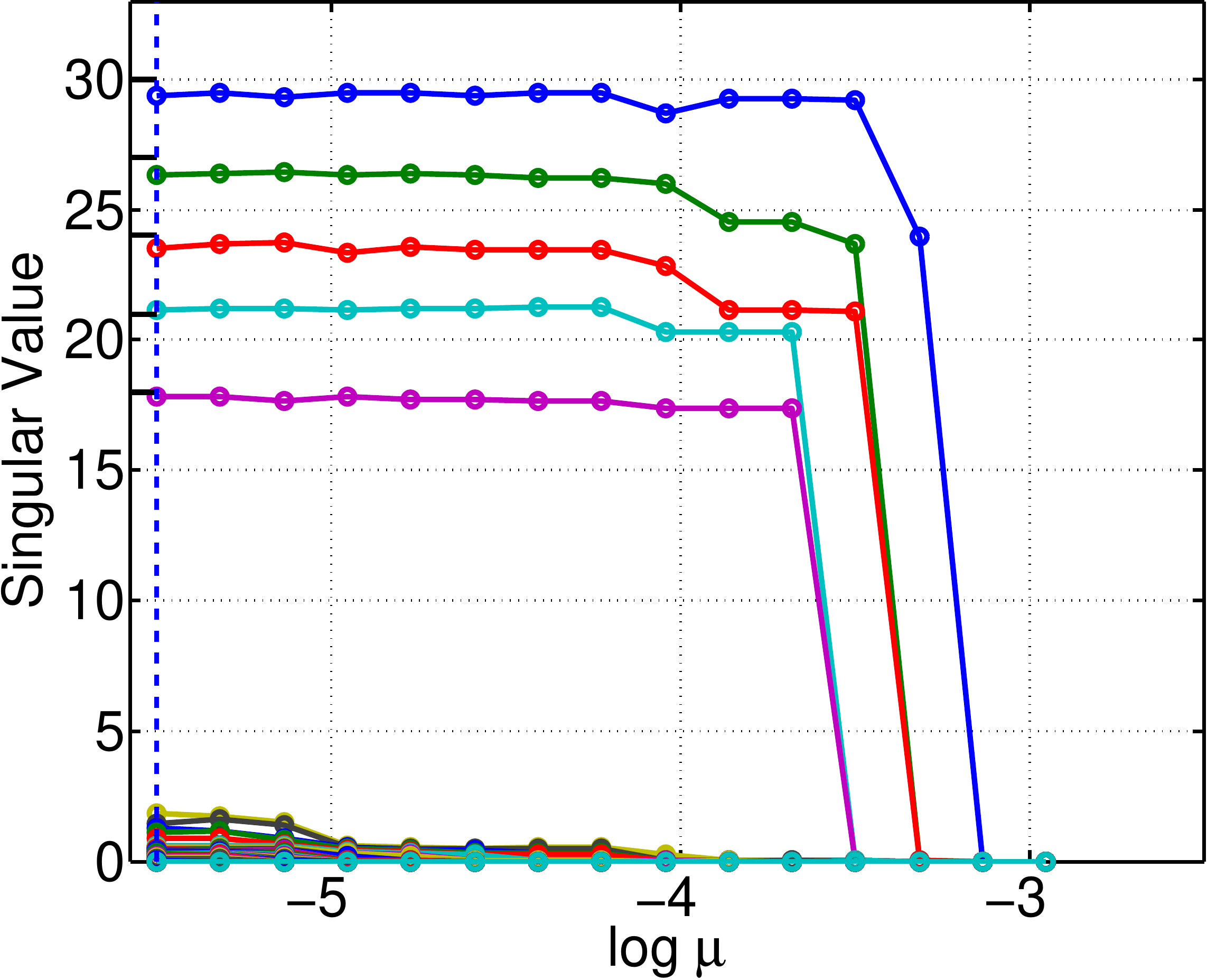}}\quad
	\subfloat[Support]{\label{fig:spSolPath}\includegraphics[scale=0.25]{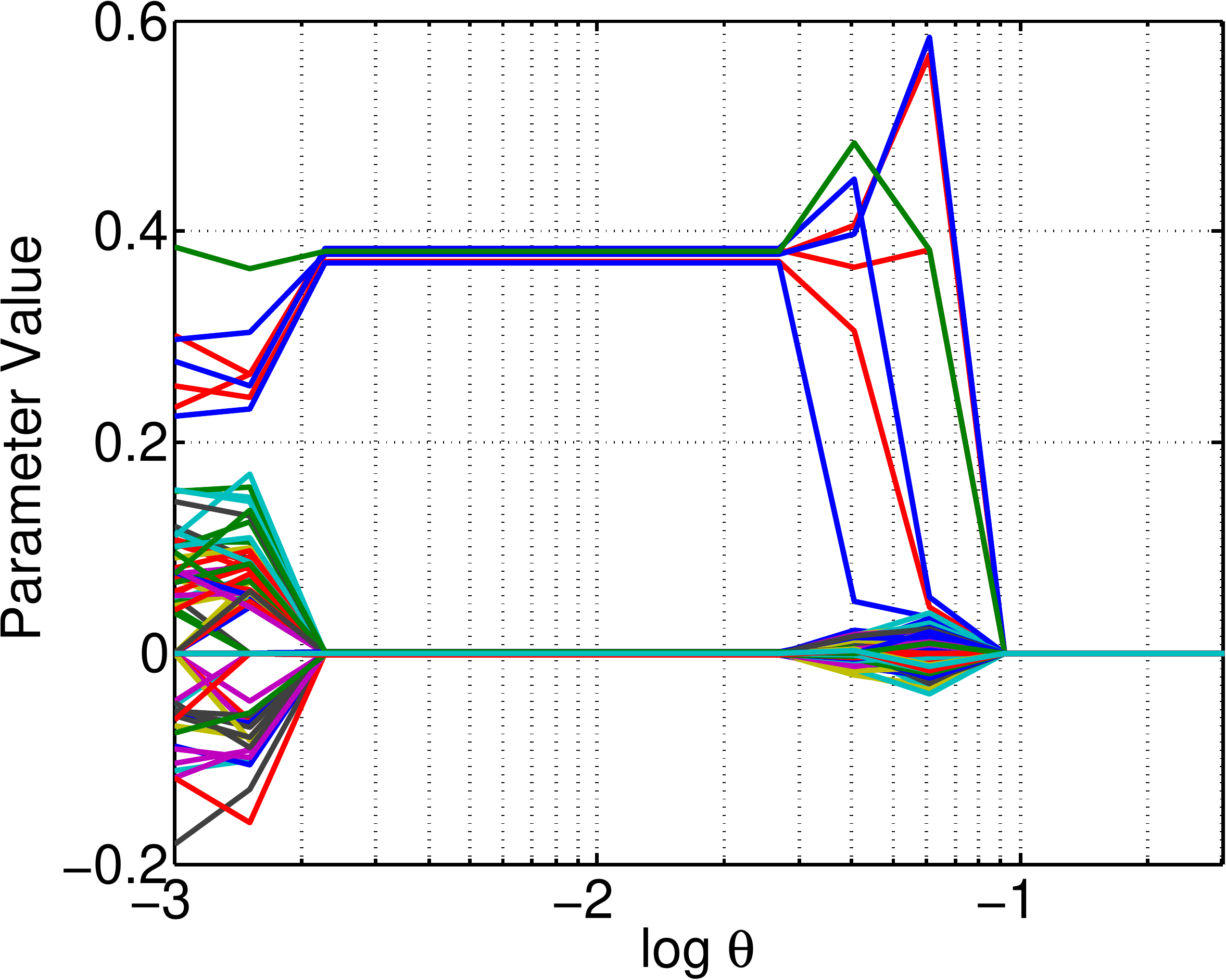}}
	\caption{(a) Solution path for the singular values of the estimated matrices. The plot show the value of top five singular values of the solution $\widehat{\bC}$ as we change the stopping error $\mu$. (b) Solution path for the top left singular vector $\mathbf{u}$ of the estimated matrices. Only seven coefficients are non-zero. The range of the parameters are generated as follows: $\mu = \mathrm{logspace}(-5, -1, 5)$ and $\theta = \mathrm{logspace}(-1, \log_{10}(20), 10)$, where $\mathrm{logspace}(a,b,n)$ indicates the minimum value $10^a$, maximum value $10^b$, and total number $n$.}
	\label{fig:solPath}
\end{figure}

Figure \ref{fig:rankPath} shows the solution path for SEED on one example data set ($p = 400$, $r=5$, $q = 200$, $\rho=0.5$, and $n = 100$).  The corresponding singular values are set to $30, 27,  24, 21$, and $18$.  In the horizontal axis, we show the termination parameter $\mu$ normalized by $\|\bY\|_F^2/(nq)$. We can see that SEED can identify the correct rank with medium values of $\mu$.
%RASE shows more stable solution path which is inline with the smaller variations of the results in Table \ref{tab:synth}.
%When $\mu$ is small, RASE makes a large number of iterations and in some of them it corrects some of its noisy steps taken in the first iterations. Thus,
Figure \ref{fig:spSolPath} shows the solution path for the top left singular vector $\mathbf{u}_1$ of $\bC$ on an example data set ($p = 200$, $q=100$, $n = 50$, $\rho=0.5$, and $r= 1$). Both of the solution paths indicate that SEED is robust to the particular choice of parameters and in a large range of parameters SEED is able to successfully recover the true rank of the matrix and the support of the singular vectors. %Only seven components of $\mathbf{u}_1^*$ are non-zero.

%%\begin{equation*}
%%\fbox{\text{Figure \ref{fig:solPath} should be about here.}}
%%\end{equation*}

\subsubsection{Simulation example 2}
In order to study scalability of SEED, we conduct the experiments on two computing environment, including: (1) an off-the-shelf personal computer (PC) and (2) a graphics processing unit (GPU), to demonstrate the runtime efficiency and the parallelization capability of SEED.

\begin{figure}[t!]
	\centering
	\subfloat[CPU ]{\label{fig:speedupCPU}\includegraphics[scale=0.25]{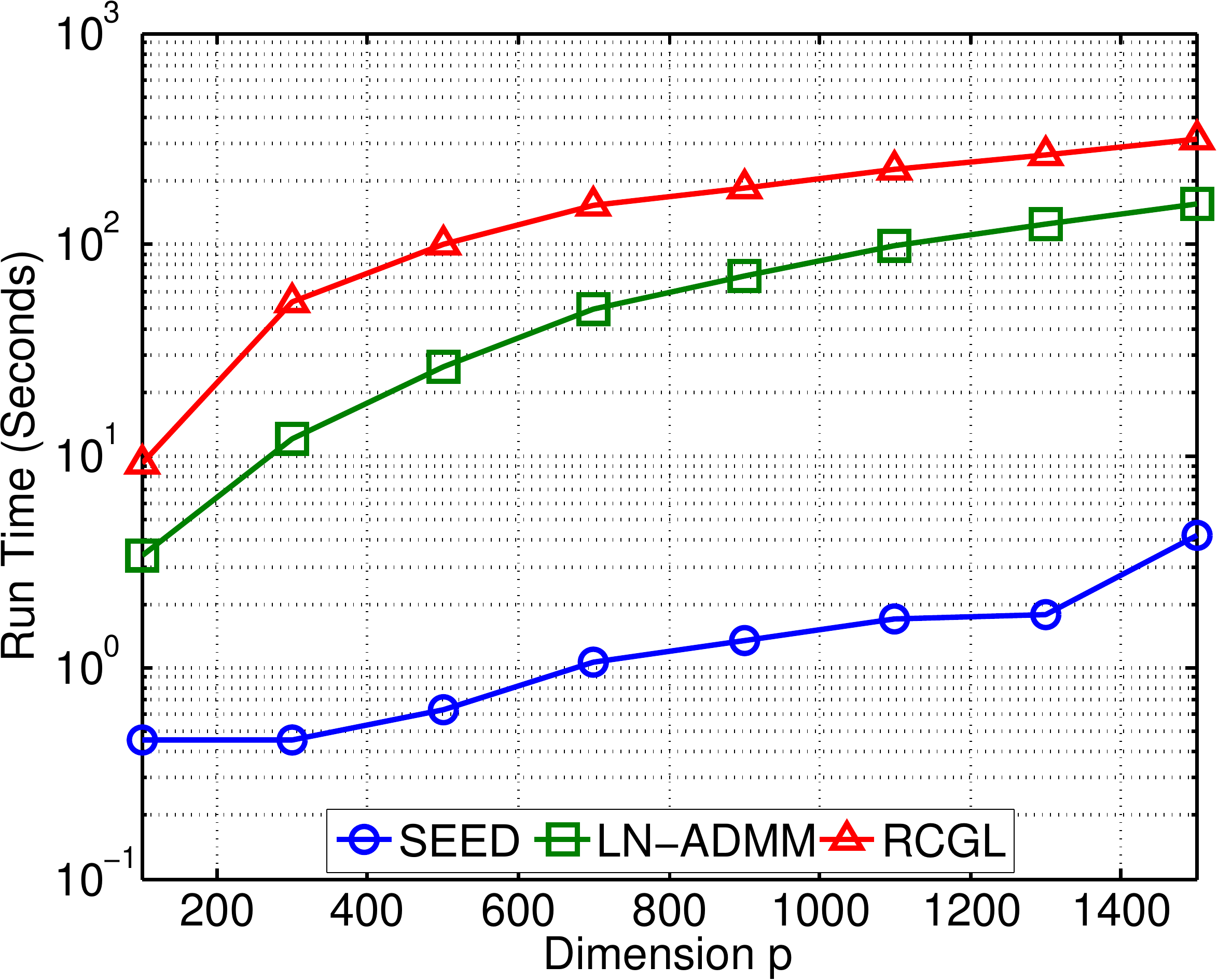}}\quad
	\subfloat[GPU ]{\label{fig:speedupGPU}\includegraphics[scale=0.25]{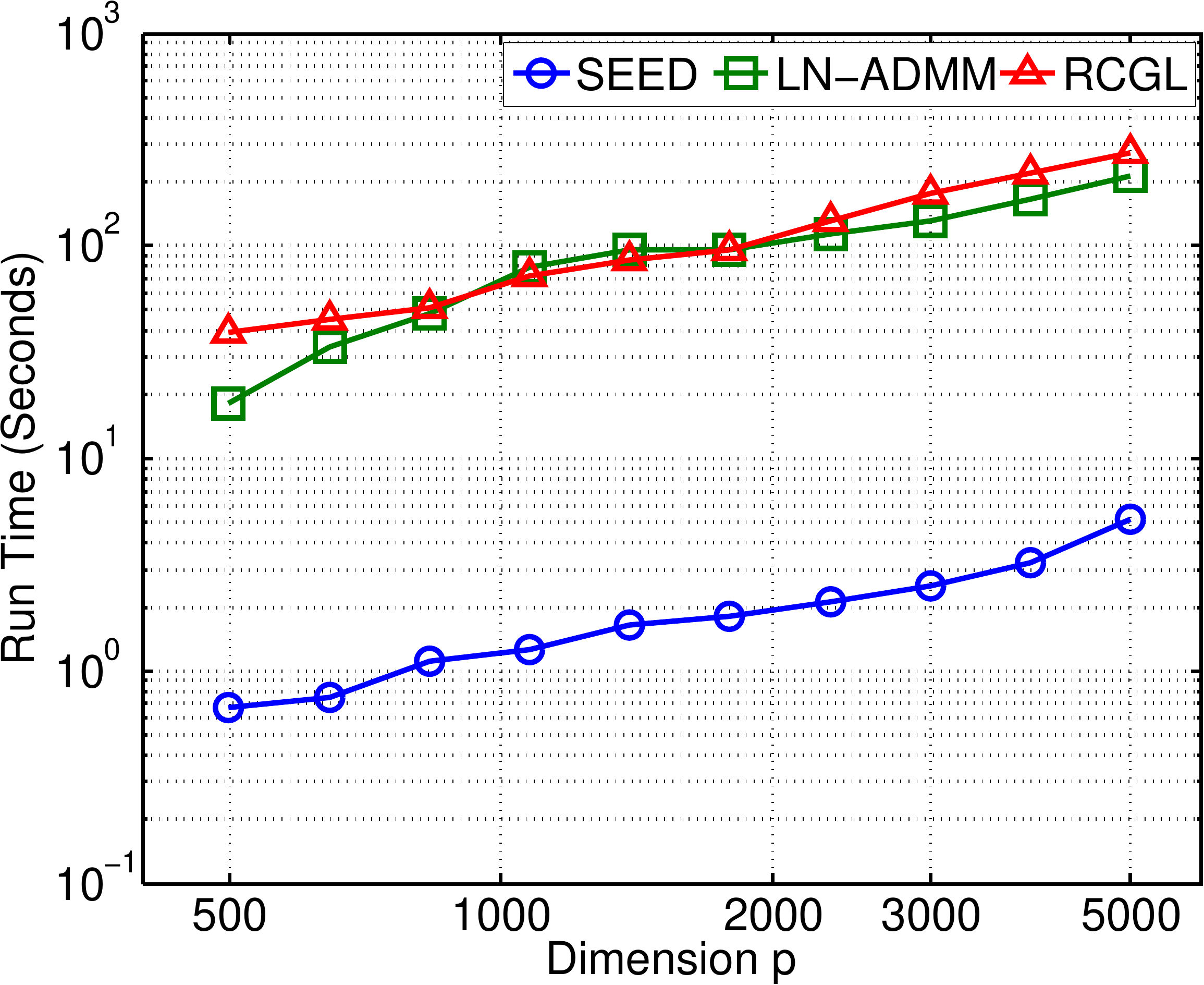}}
	\caption{Speedup by SEED on (a) CPU and (b) GPU devices.  Note that the vertical axis is in logarithmic scale.}
	\label{fig:speedup}
\end{figure}

First, we run our experiments on an off-the-shelf PC with Intel i7 at 3.4GHz and 8GB of memory. The system runs MATLAB R2013b on the Windows operating system. We generate 5 data sets with $r = 1$, non-zero ratio of $10\%$, $q = 1000$, and $n = 1000$. Figure \ref{fig:speedupCPU} shows the average CPU runtime of three algorithms as the dimension $p$ increases. We can see that SEED can achieve a speed up of  10-100 times in runtime compared with baseline methods.

%%\begin{equation*}
%%\fbox{\text{Figure \ref{fig:speedup} should be about here.}}
%%\end{equation*}
%%

\begin{figure}[t!]
	\centering
	\subfloat[ ]{\label{fig:gTime1}\includegraphics[scale=0.25]{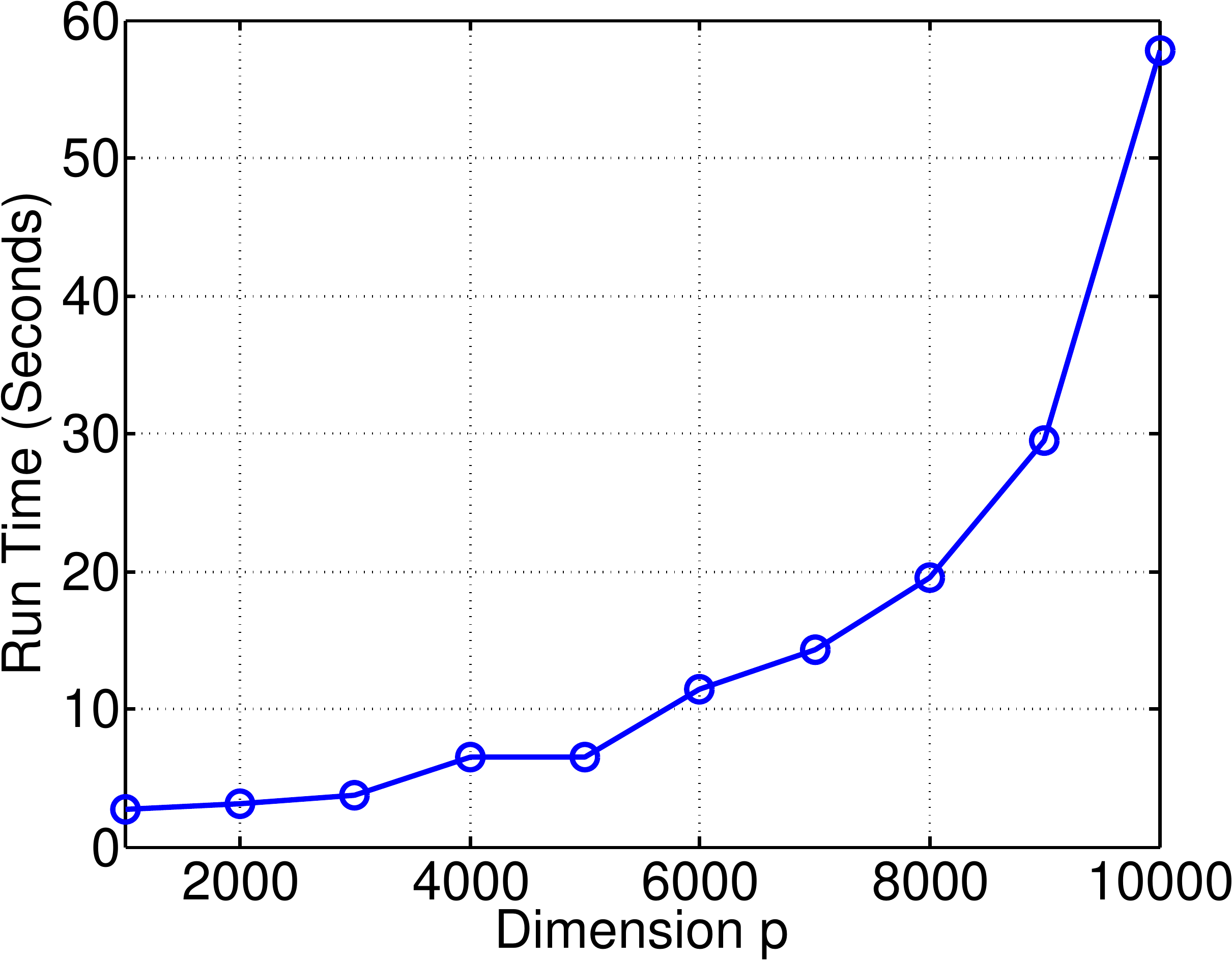}}~
	\subfloat[ ]{\label{fig:gPar1}\includegraphics[scale=0.25]{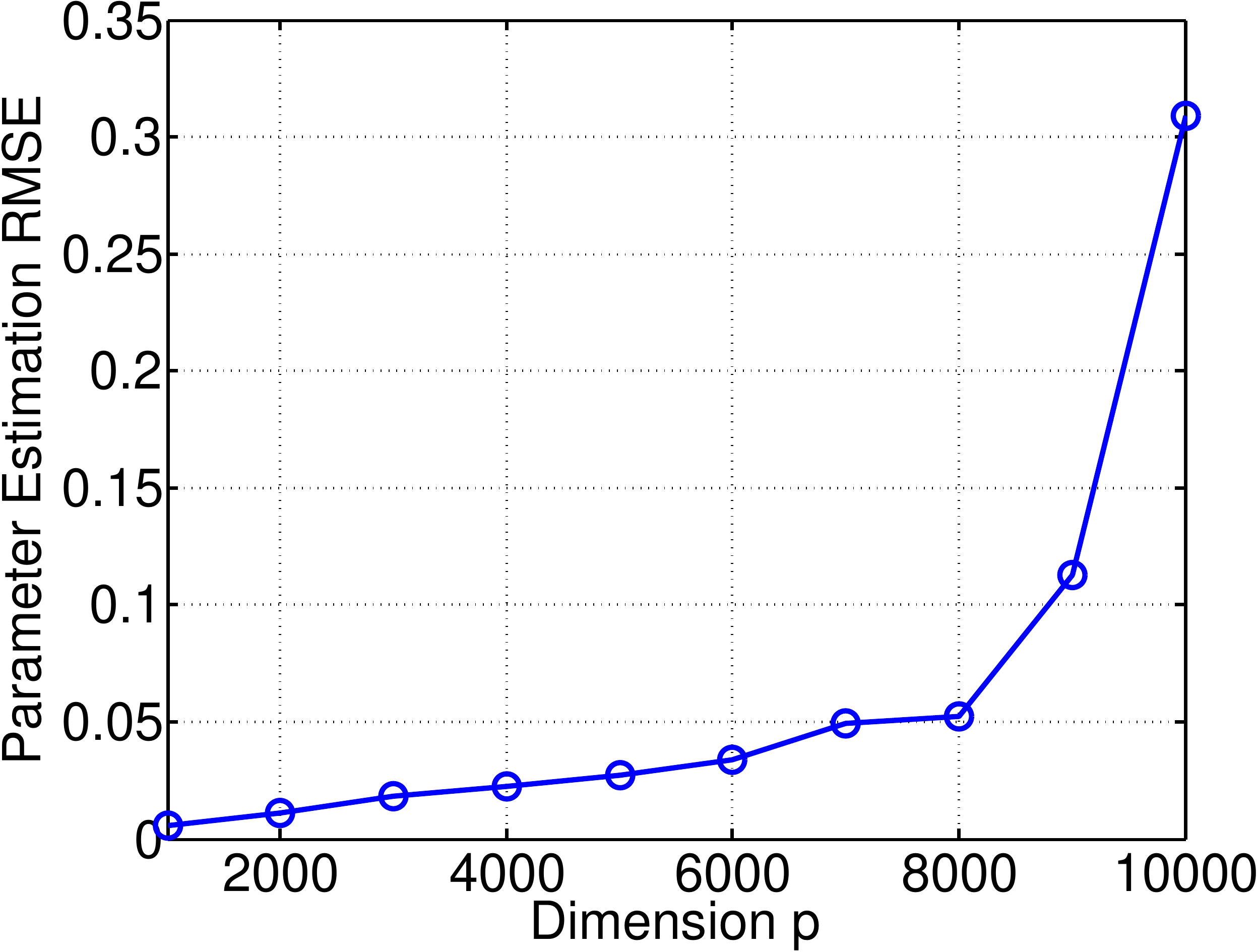}}
	\caption{Scalability experiments on very large data sets on GPU using the fast approach. Accuracy results are normalized.}
	\label{fig:scalability}
\end{figure}

Next, in order to test scalability of SEED in extremely large data sets, we use a machine that is equipped with a Tesla K40 GPU which has 2880 processing cores at 745MHz and 12GB of memory. We perform our experiments with MATLAB R2013b on a Debian Linux operating system. GPUs are built to have many less-powerful processing units which makes them ideal for parallel implementation \cite[Chapter 5]{bekkerman2012}. %Our experiments on GPU are indented to demonstrate that RASE can be efficiently parallelized and therefore practical for large-scale problems.
Therefore we apply the two-step fast eigenvalue decomposition described in Section \ref{sec:imp}, which involves only simple matrix operation and can be paralleled easily. %Figure \ref{fig:scalability} shows the GPU runtime of different algorithms on an example large-scale data set ($q = 10000$, $n = 5000$, $r = 1$ and non-zero ratio of $10\%$).
The experiment results shown in Figures \ref{fig:speedupGPU} and \ref{fig:scalability} are obtained under the setting of $q = 10000$, $n = 5000$, $r = 1$ and non-zero ratio of $10\%$. The results indicate that while SEED is fast on the GPU, it also achieves reasonable accuracy. Note that the results show that SEED is able to estimate a sparse and low-rank matrix with $10^8$ elements in less than a minute, confirming its extreme scalability.
% We can see that the GPU solution of RASE provides near 10-time speedup over PC while it still achieves reasonable accuracy.
%%\begin{equation*}
%%\fbox{\text{Figure \ref{fig:scalability} should be about here.}}
%%\end{equation*}

%Note that the limiting factor in the experiments with GPU is the
%\begin{figure}[t!]
%\centering
%\label{fig:rmseB}\includegraphics[scale=0.35]{./Figs/gpuOnly}
%\caption{The log-log plot of the runtime to demonstrate empirical sub-linear scalability of RASE on GPU.}
%\label{fig:huge}
%\end{figure}

%Finally, the log-log plot of run time vs dimensionality in Figure \ref{fig:huge} shows that under parallel implementation, RASE can scale linearly in terms of dimension. The largest parameter matrix in Figure \ref{fig:huge} has $p = 16,000$ and $q = 1000$ and the algorithm terminated on average in $7.0$ seconds.

\subsection{Real data analysis}
\textit{Diffusion Network Inference}, that is, the task of inferring influence networks from user activities, is one of the central tasks in social networks analysis \citep{gomez2012} because it helps improve social marketing by finding the influential users in a network.  It is a challenging problem because: (i) in many social networks the influence is expressed implicitly \citep{gomez2012} and (ii) empirical studies show that common metrics such as number of friends or followers fail to accurately measure the social influence of the users \citep{cha2010}.

A popular computational approach in estimating social influence among users is to count the number of users' activities over a time span (in regularly or irregulary spaced intervals) and analyze the resulting time series data \citep{Truccolo2005}. Many different models have been developed, among which the vector auto-regressive model arises as a simple and robust solution \citep{trusov2009effects,Bahadori2013}. That is, every user is described by a time series $x_i(t)$ for $t = 1, \ldots, T$. Next, the vector auto-regressive model with $L$ lags is fitted to the time series as follows:
\begin{equation*}
\mathbf{x}(t) = \sum_{\ell = 1}^{L} A_{\ell}\mathbf{x}(t-\ell),
\label{eq:var}
\end{equation*}
where $\mathbf{x}(t) = [x_1(t), \ldots, x_p(t)]$ and $A_{\ell}$ is the evolution matrix at $\ell$th lag. The influence network can be built from the evolution matrices by establishing an edge from node $j$ to node $i$ if $\sum_{\ell = 1}^{L} |A_{\ell, i, j}|$ is significantly larger than zero.

In this experiment, we gather a Twitter data set with tweets on the ``Haiti earthquake'' and apply vector auto-regressive model to identify the potential top influencers on this topic (that is, those Twitter accounts with the largest impact on the others). We divide the 17 days after the Haiti Earthquake on Jan. 12, 2010 into 1000 uniformly spaced intervals and generate a multivariate time series data set by counting the number of tweets on this topic for the top 1000 users who tweeted most about it.
%The resulting time series have on average $0.0225$ tweets per user per bin which shows how infrequent the activities in the data set are.
For accurate modeling, we remove the users that were highly correlated with each other, most of which were operated by the same users and tweeted exactly the same content. We also remove robot-like user-accounts which tweeted on very regular intervals, which in total led to a subset of 270 users. We analyze this data with a VAR($5$) model which requires estimation of a $q = 270$ dimensional response vector using $p = 1350$ predictors while we have only $n= 995$ observations. The number of lags is chosen based on the intuition about the maximum retweeting delay.

Since we do not have access to the true influence network, we use the retweet network as a surrogate of the ground truth following the evaluation convention in the social networks community. The retweet network is constructed by adding an edge from user $i$ to user $j$ if user $j$ has retweeted at least $4$ of the tweets of user $i$. Clearly, the retweet network is not the actual underlying temporal dependency graph, mainly because there are possible implicit influence patterns as well. However, it is the best possible metric that we could obtain for graph estimation accuracy evaluation in our data set \citep{cha2010}. The retweet network for the 270 selected users is sparse; it has only $0.11\%$ of possible edges.% and low-rank as well; the Ky Fan-5 norm of the retweet network is $0.5844$ of its nuclear norm.

\begin{figure}[t!]
	\centering
	\includegraphics[scale=0.25]{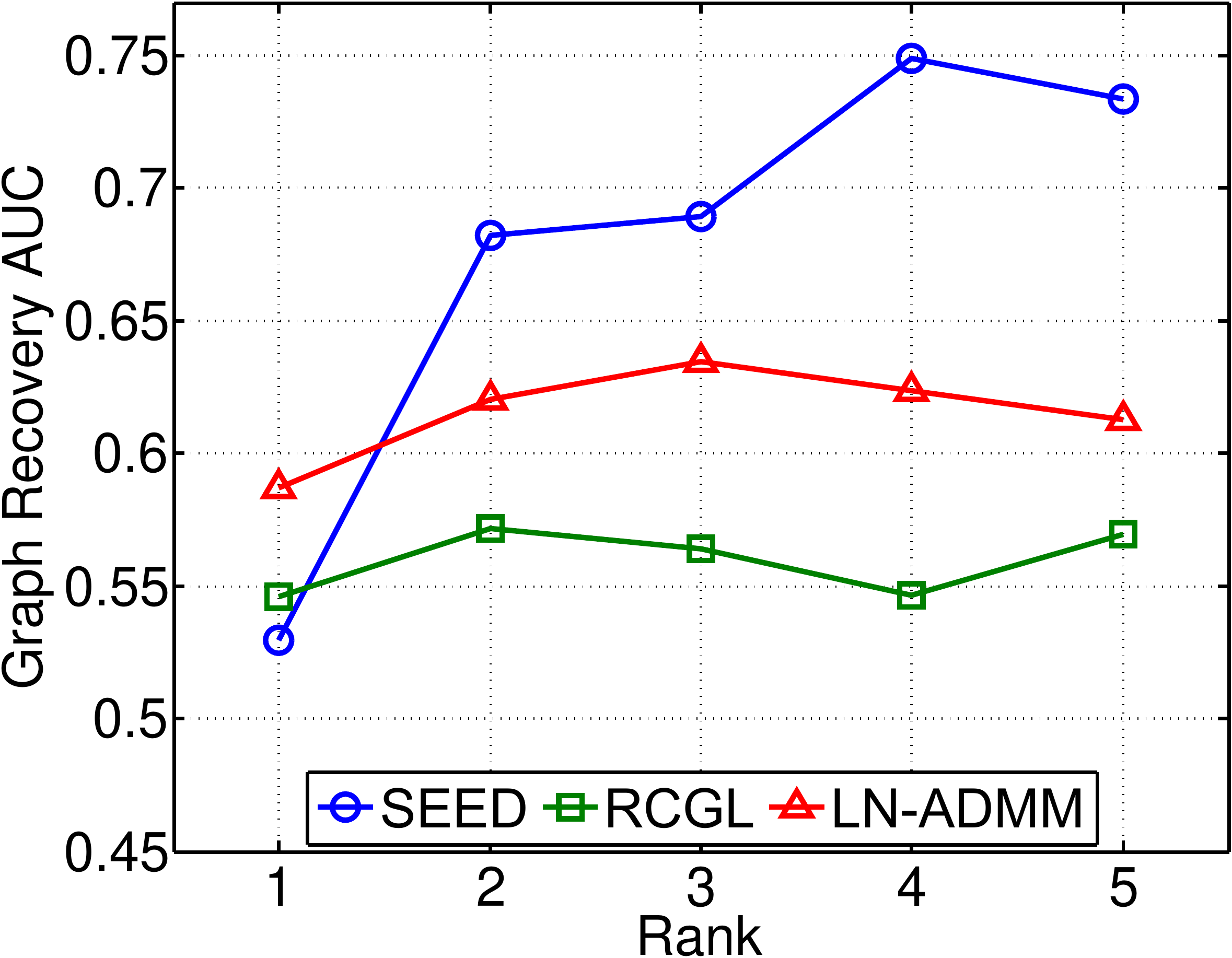}
	\caption{The graph recovery accuracy of the algorithms as the rank of solution varies. }
	\label{fig:accuracyTwitter}
\end{figure}

%It has been suggested that the influence network should have simultaneous sparse and low-rank structure \cite{Zhou2013}.
We apply SEED, LN--ADMM, and RCGL algorithms to uncover the influence network in our twitter data set. Figure \ref{fig:accuracyTwitter} shows the accuracy of the procedures in uncovering the true influence network in terms of AUC.
For every value of the rank parameter, we tune the sparsity by 5-fold cross-validation. Given the fact that exact rank constraint cannot be enforced directly in the LN--ADMM
 algorithm,  we find the best value of the nuclear norm regularization parameter $\lambda_L$ by 5-fold cross-validation. Then, we compute the low-rank approximations of the parameter matrix and evaluate the accuracy at each rank.
%%\begin{equation*}
%%\fbox{\text{Figure \ref{fig:accuracyTwitter} should be about here.}}
%%\end{equation*}

The results in Figure \ref{fig:accuracyTwitter} show that SEED significantly outperforms the baseline procedures. They also indicate that, in all of the algorithms, as we increase the rank of the solution matrix, the accuracy is improved initially and then quickly saturates. SEED achieves the highest accuracy when the rank is 4. Note that this result also confirms other studies that the social network connections may be strongly influenced by a few unobserved exogenous variables \citep{myers2012information}. The results in Table \ref{tab:speedTwitter} demonstrate the significant speedup achieved by SEED compared to the baselines.
%%%%\begin{equation*}
%%%%\fbox{\text{Table \ref{tab:speedTwitter} should be about here.}}
%%%%\end{equation*}

\begin{table}[t]
	\caption{Run time (in seconds) of the algorithms on the application data set. }
	\centering
	\begin{tabular}{ c|c|c }
		\toprule
		{SEED} &{LN--ADMM
		} &{RCGL}\\
		\toprule
		2.83 &127.34 & 256.87\\
		\bottomrule
	\end{tabular}
	\label{tab:speedTwitter}
\end{table}

\section{Discussion}
\label{sec:dis}

In this paper, we propose to convert the problem of sparse reduced-rank regression into a sparse generalized eigenvalue problem, which allows us to efficiently employ the recently developed sparse eigenvalue decomposition techniques. After this transformation, the left singular vectors can be estimated in two simple steps, and the estimation of both sparse and dense right singular vectors is unified in a single framework. As a pure learning algorithm, SEED deviates from traditional regularization frameworks (i.e., a loss function plus certain penalties), leading to computational efficiency and scalability. Furthermore, we prove that SEED achieves nice estimation and prediction accuracy that coincides with the minimax error bound in the univariate regression setting \citep{yu2011}.

Some interesting problems for future research include extending the current formulation of the regression coefficient matrix in (\ref{eq:C*}) to the case where the singular values can be repeated such that the left singular vectors (which correspond to latent factors) are not identifiable. Then we will need to estimate the eigenspaces spanned by important singular vectors and characterize the estimation accuracy by some new criterion, such as the one in \cite{cai2013} and \cite{Ma2013}. Another research direction is to explore the theory of random design matrices and this can be addressed by using an extended version of perturbation theory where the perturbation in $\bP$ is also included in the analysis.

Moreover, it is computationally straightforward to extend SEED to the generalized linear models by adapting the sequential quadratic programming framework. For this extension, we first approximate the loss function by the quadratic loss function and find the optimal unit rank matrix. Then we can add the unit rank matrix to the solution and re-approximate the loss function with another quadratic function around this new solution. By performing these three steps sequentially, we can efficiently estimate the low-rank coefficient matrix. Statistical properties of such estimator can be analyzed by extending the results in \cite{lozano2011group} for greedy sparse procedures to reduced-rank regression.

\bibliographystyle{apalike}
%\bibliography{vSelect}

\end{document}